\documentclass[12pt]{article}

\usepackage{scicite}

\usepackage{times}
\usepackage{url}

\newenvironment{sciabstract}{%
\begin{quote} \bf}
{\end{quote}}

\usepackage{graphicx}
\graphicspath{{Img/}}
\usepackage{caption}
\usepackage{graphicx}
\usepackage{subfigure} 
\usepackage{amsmath}
\usepackage{array}
\graphicspath{{Img/}}
\usepackage{caption}
\usepackage{threeparttable}
\usepackage{overpic}

\title{Brain-inspired bodily self-perception model for robot rubber hand illusion}

\author
{Yuxuan Zhao$^{1}$, Enmeng Lu$^{1}$, Yi Zeng$^{1,2,3,4,5\star}$\\
\\
\normalsize{$^{1}$Brain-inspired Cognitive Intelligence Lab, Institute of Automation,}\\
\normalsize{Chinese Academy of Sciences, Beijing, China.}\\
\normalsize{$^{2}$State Key Laboratory of Multimodal Artificial Intelligence Systems,}\\
\normalsize{Chinese Academy of Sciences, Beijing, China.}\\
\normalsize{$^{3}$Center for Excellence in Brain Science and Intelligence Technology,}\\
\normalsize{Chinese Academy of Sciences, Shanghai, China.}\\
\normalsize{$^{4}$University of Chinese Academy of Sciences, Beijing, China.}\\
\normalsize{$^{5}$Center for Long-term Artificial Intelligence, Beijing, China.}\\
\\
\normalsize{To whom correspondence should be addressed; E-mail:  yi.zeng@ia.ac.cn.}
}

\date{}

\begin{document} 


\baselineskip24pt


\maketitle


\begin{sciabstract}
At the core of bodily self-consciousness is the perception of the ownership of one's body. Recent efforts to gain a deeper understanding of the mechanisms behind the brain's encoding of the self-body have led to various attempts to develop a unified theoretical framework to explain related behavioral and neurophysiological phenomena. A central question to be explained is how body illusions such as the rubber hand illusion actually occur. Despite the conceptual descriptions of the mechanisms of bodily self-consciousness and the possible relevant brain areas, the existing theoretical models still lack an explanation of the computational mechanisms by which the brain encodes the perception of one's body and how our subjectively perceived body illusions can be generated by neural networks. Here we integrate the biological findings of bodily self-consciousness to propose a Brain-inspired bodily self-perception model, by which perceptions of bodily self can be autonomously constructed without any supervision signals. We successfully validated our computational model with six rubber hand illusion experiments and a disability experiment on platforms including a iCub humanoid robot and simulated environments. The experimental results show that our model can not only well replicate the behavioral and neural data of monkeys in biological experiments, but also reasonably explain the causes and results of the rubber hand illusion from the neuronal level due to advantages in biological interpretability, thus contributing to the revealing of the computational and neural mechanisms underlying the occurrence of the rubber hand illusion.

\end{sciabstract}


\section{Introduction}
At the core of bodily self-consciousness is the perception of the ownership of one's own body \cite{RN604,RN981}. In recent years, in order to understand more deeply how the brain encodes self body, some researchers have tried to establish a unified theoretical framework to explain the behavioral and neurophysiological phenomena in the coding of bodily self-consciousness from the perspectives of Predictive coding and Bayesian causal inference. For these theoretical models, a core problem that needs to be explained is how body illusions such as the rubber hand illusion occur. The rubber hand illusion refers to the illusion that occurs when the participant's rubber hand and the invisible real hand are presented with synchronized tactile and visual stimuli, in which the participant will have the illusion that the rubber hand seems to become his or her own hand \cite{RN993}.

Prediction coding method realizes body estimation by minimizing the errors between perception and prediction\cite{RN994}. Hinz et al. \cite{RN983} regard the rubber hand illusion as a body estimation problem. They take humans and a multisensory robotic that can perceive the proprioception, vision and tactile information on its arm as participants, and verify their model through the traditional rubber hand illusion experiment. The experiment shows that the proprioception drifts are caused by prediction error fusion instead of hypothesis selection. 

The Active inference models can be regarded as an extension of Predictive coding approaches. Rood et al. \cite{RN990} propose a deep active inference model for studying the rubber hand illusion. They model the rubber hand illusion experiment in a simulated environment. The results show that their model produces similar perceptual and active patterns to that of human during the experiment. Maselli et al. \cite{RN986} propose an active inference model for arm perception and control, which integrates the imperatives both in intentional and conflict-resolution. The imperative of intentional is used to control the achievement of external goals, and the imperative of conflict-resolution is to avoid multisensory inconsistencies. The results reveal that intentional and conflict-resolution imperatives are driven by different prediction errors.

Bayesian causal inference model is extensively used in the theoretical modeling of multimodal integration, and has been repeatedly verified at the behavioral and neuronal levels \cite{RN639,RN988}. The Bayesian causal inference model can well reproduce and explain a variety of rubber hand illusion experiments. Samad et al. \cite{RN970} adopted the Bayesian causal inference model of multisensory perception, and test it through the rubber hand illusion on humans. The results show that their model could reproduce the rubber hand illusion, and suggest that the Bayesian causal inference dominates the perception of body ownership. Fang et al. \cite{RN580} present a nonhuman primate version experiment of a rubber hand illusion. They conduct the experiment on human and monkey participants. They adopted the Bayesian causal inference (BCI) model to establish an objective and quantitative model for the body ownership of macaques. With the BCI model, they investigate the computational and neural mechanisms of body ownership in the macaque brain from monkey behavioral and neural data, respectively. The results show that the BCI model can fit monkey behavioral and neuronal data well, helping to reveal a cortical representation of body ownership.

Although the above theoretical models describe the mechanism of bodily self-consciousness and possible related brain areas from a conceptual perspective, they still do not explain the computational mechanism of how the brain encodes bodily self-consciousness and how the body illusion we subjectively perceive is generated by neural networks.

Here we integrate the biological findings of bodily self-consciousness and construct a Brain-inspired bodily self-perception model, which can achieve association learning of proprioceptive and visual information and construct the bodily self-perception model autonomously without any supervision signals. In terms of experiments, this model can reproduce six rubber hand illusion experiments simultaneously at behavioral and neuronal scales, and the experimental results can well fit the behavioral and neural data of monkeys in biological experiments. In terms of biological explainability, this model can explain the causes and results of the rubber hand illusion reasonably from the neuron scale, which is helpful to reveal the computational and neural mechanism underlying the rubber hand illusion. In addition, the results of the disability experiment indicate that the generation of rubber hand illusion is the result of the joint action of primary multisensory integration area (e.g., TPJ) and high-level multisensory integration area (e.g., AI), and neither is indispensable.

\section{Results}

In this section, we first introduce the Brain-inspired bodily self-perception model, then introduce the design of the experiment, and finally introduce the Proprioceptive drift experiment result on the iCub humanoid robot, and the results of Proprioceptive drift, Proprioception accuracy, Appearance replacement, Asynchronous, Proprioception only, Vision only and Disability experiments in the simulated environment.

\subsection{Brain-inspired bodily self-perception model}

\begin{figure}[htbp]
\centering
\subfigure[Brain-inspired Bodily Self-perception Model]{
\includegraphics[width=2.7in]{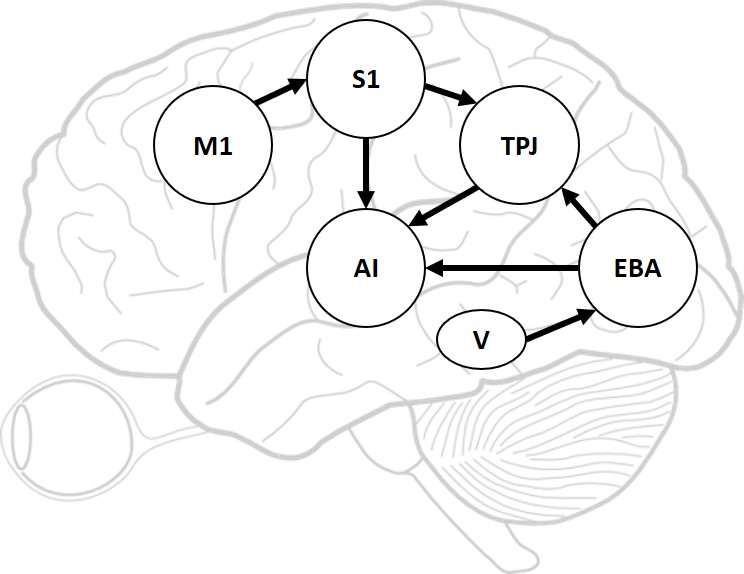}
}
\quad
\subfigure[Neural network architecture of the model]{
\includegraphics[width=2.7in]{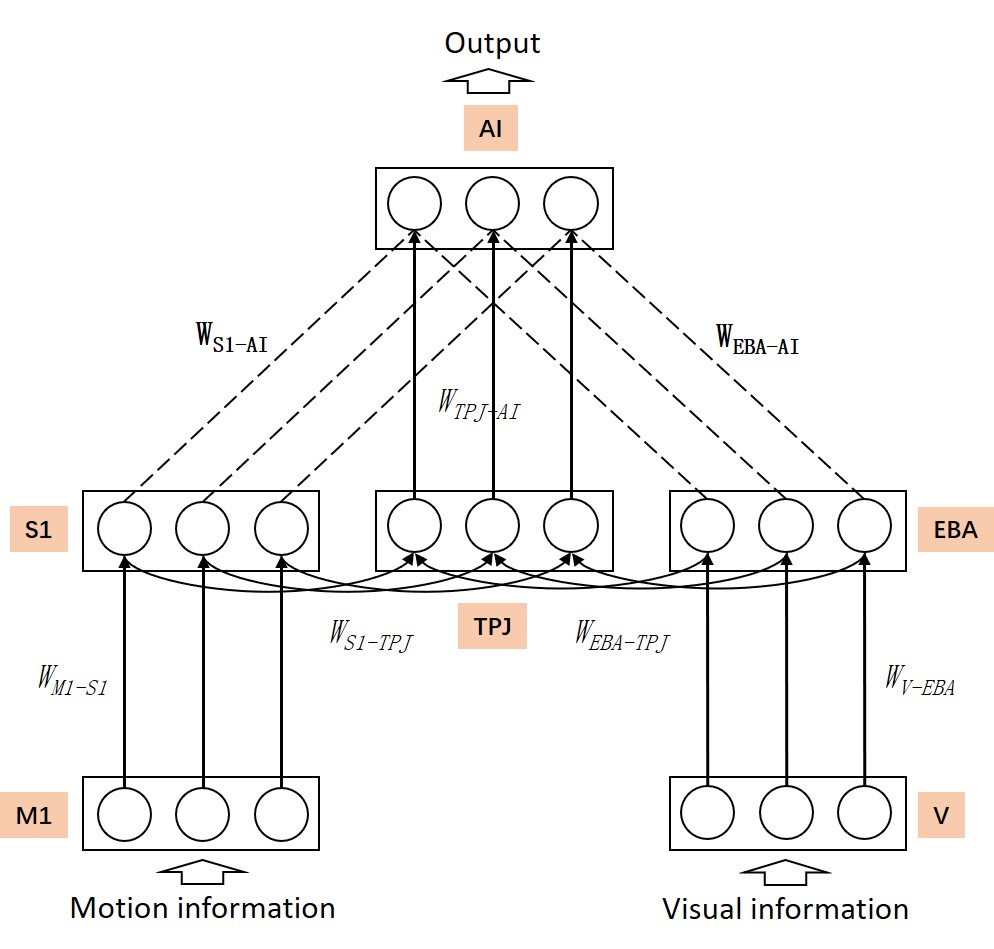}
}
\caption{ The architecture of the Brain-inspired Bodily Self-perception Model and its neural network architecture. \textbf{(a)} Brain-inspired Bodily Self-perception Model (including major functional brain areas, pathways, and their interactions), and the arrow indicates the direction of information transmission. M1, Primary motor cortex; S1, Primary somatosensory cortex; TPJ, Temporo-parietal junction; V, Vision input; EBA, Extrastriate body area; AI, Anterior insula.
\textbf{(b)} The neural network architecture of the model. Brain areas are shown in orange, neurons in brain areas are shown in circles, and the connections between different brain areas are shown in black lines. The dotted lines denote excitatory or inhibitory synapses which are depended on the results of synaptic plasticity (such as $W_{S1-AI} and W_{EBA-AI}$ in bold), and the solid lines indicate fixed synaptic weights (such as $W_{M1-S1}, W_{S1-TPJ}, W_{TPJ-AI}, W_{V-EBA},  W_{EBA-TPJ}$ in italics).}
\label{model}
\end{figure}

We integrate the biological findings of bodily self-perception into a model to construct a brain-inspired model for robot bodily self-perception, and the architecture of the model is shown in Figure \ref{model}A. 

We have derived and designed our computational model to include multiple functional brain areas based on the current biological understanding of the functions of various brain areas involved in bodily self-perception.
Specifically, the primary motor cortex (M1) is considered for encoding motion's strength and orientation, and controlling the motion execution \cite{RN251,RN997}. In our computational model, the M1 is used to encode the orientation of motion. 
The primary somatosensory cortex (S1) is considered for proprioception to perceive limb position \cite{RN995,RN996}. In our computational model, S1 receives stimuli from M1 for the perception of arm movement orientation.
The extrastriate body area (EBA) is considered to be involved in the visual perception of the human body and body parts \cite{RN974}, and in the rubber hand illusion experiment, it is proved to be directly involved in the processing of limb ownership \cite{RN972}. In our computational model, the EBA receives information from Vision (V) and obtains finger orientation information.
The temporo-parietal junction (TPJ) integrates information from visual, auditory, and somatosensory systems and is considered to play an important role in multisensory bodily processing \cite{RN975,RN605}, and has been reported be activated in the rubber hand illusion experiment \cite{RN1003,RN1004}. In our computational model, the TPJ receives information from S1 and EBA, and performs primary multisensory information integration.
The anterior insula (AI) is considered to be a primary region for processing interoceptive signals \cite{RN604}, multimodal sensory processing \cite{RN1000,RN998,RN999}, and it has also been reported to be activated in the rubber hand illusion experiment \cite{RN1002,RN1003}. In our computational model, the AI receives information from S1, TPJ and EBA, performs high-level multisensory information integration, and outputs the result.

Figure \ref{model}B shows the neural network architecture of the proposed model. This model is a three-layer spiking neural network model with cross-layer connections. The M1 area generates information about the robot's motion direction and the V area receives visual information about the robot. The AI area outputs the robot's behavior decision information according to the highest firing rate of the neuron. Unsupervised learning of the bodily self-perception model is realized through the motor and visual information, and tested using the rubber hand illusion adapted from the macaque behavior experiment \cite{RN580}. More details could be found in Materials and Methods section.

\subsection{Experiments on the iCub robot}

\subsubsection{Experimental design}

\begin{figure}[htbp]
\centering
\includegraphics[width=15cm]{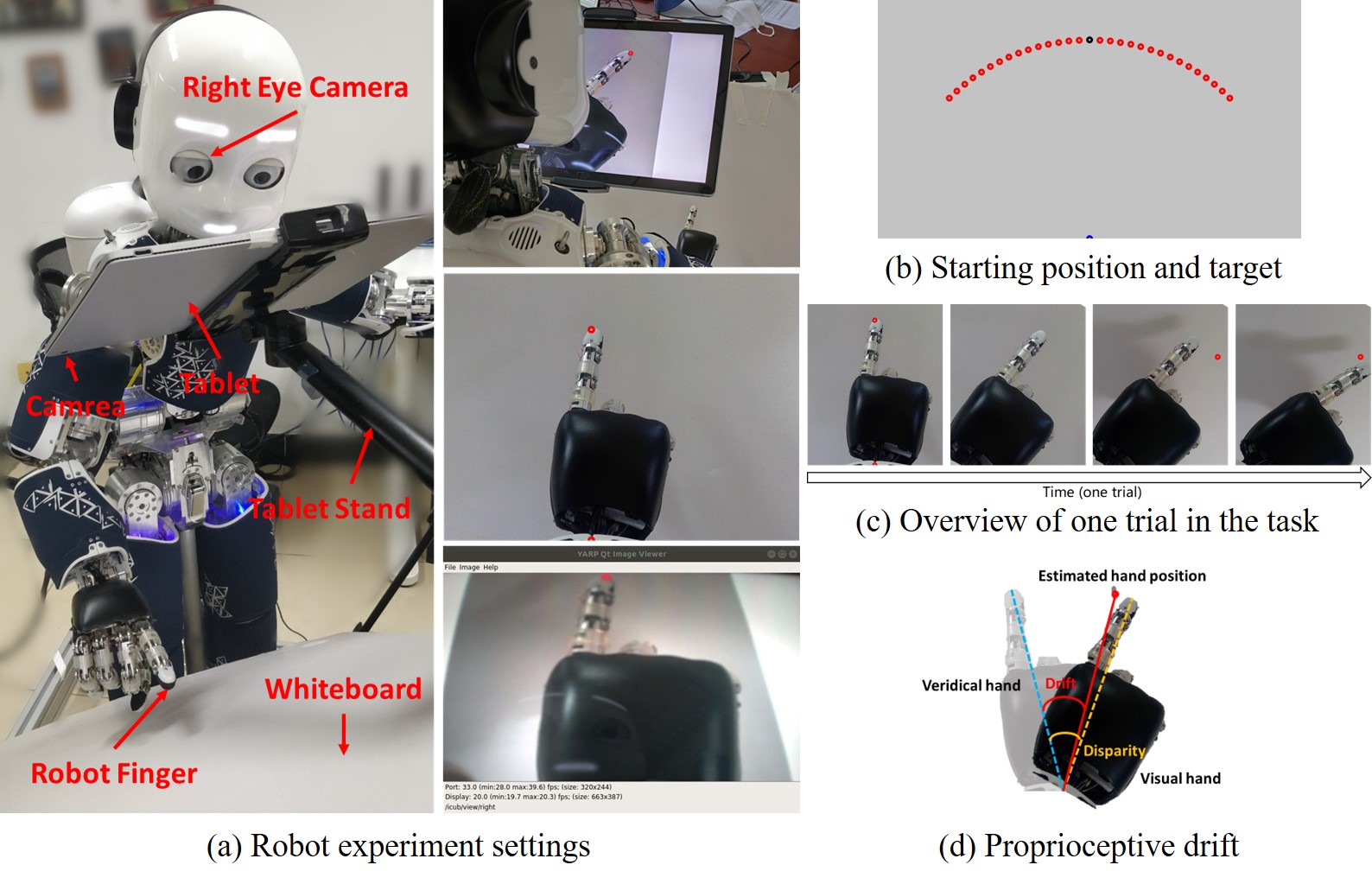}
\caption{Overview of the robot experiment. (a) Robot experiment settings. 
(b) Starting position and target. 
(c) Overview of one trial in the task.
(d) Proprioceptive drift. }
\label{RobotE}
\end{figure}

The overview of the robot experiment is shown in Figure \ref{RobotE}. 

Figure \ref{RobotE}A shows the scene settings of the experiment. The left panel shows the equipment involved in the robot experiment, including an iCub humanoid robot, a Surface Book 2 tablet with a camera in its back, a tablet stand, and a white cardboard. The robot's hand is placed in between the tablet and the white cardboard. The tablet serves two functions: blocking the robot from seeing its own hand directly; capturing images of the robot's hand through the back camera, rotating the image, and displaying the processed image in front of the robot's own eye-camera (the right eye-camera). The tablet stand is used to support and hold the tablet in a desired position. The white cardboard blocks sundries to make the background as clean as possible, thus reducing the difficulty of image recognition and the possibility of incorporating other intrusive visual cues. The settings of the equipment make sure that the robot could only perceives its visual hand through the display of the tablet by its own eye-camera. The right side panel shows, from top to bottom, a scene of the robot experiment in the right rear view, the image displayed by the tablet and the image perceived by the robot using its right eye-camera. 

Figure \ref{RobotE}B shows the Starting position and target during the robot experiment. All points in the experiment are colored in red, and here we have marked the initial positions as blue and black points for ease of illustration. The robot first places its wrist on the blue dot below the image and places its finger on the black dot. The red dots are the target that the robot needs to point to during the experiment. These targets are distributed on a circular arc with the blue dot as the circular dot (the position of the robot wrist) and the distance between the blue dot and the black dot as the radius (the distance of the robot's wrist to its fingertip). With the point directly in front of the circular dot as $0^{\circ}$, ranging from $-45^{\circ}$ to $45^{\circ}$, and the interval is $15^{\circ}$. There are 9 target points in total.

Figure \ref{RobotE}C shows the overview of one trial in the task. (1) The robot puts its hand in the starting position. (2) The point at the starting position disappears, and the visual hand rotates at a random angle. Considering the motion range and precision of the robot, the rotation angle ranges from $-36^{\circ}$ to $36^{\circ}$, and the interval is $6^{\circ}$, including two angles $-45^{\circ}$ and $45^{\circ}$. There are 15 rotation angles in total. (3) The target is displayed. (4) Based on multisensory integration by vision and proprioception, the robot makes behavioral decisions and points to the target.

Figure \ref{RobotE}D shows the measurement method of proprioceptive drift. The dark hand is the visual hand, and the light hand is the veridical hand. The red target point is the position of the hand estimated by the robot, while the veridical hand is the position where the proprioceptive hand is located. Therefore, the angle between the robot's veridical hand and the target point is the proprioceptive drift in the rubber hand illusion, that is, the angle between the blue dotted line and the red solid line. The angle between the blue dotted line and the orange dotted line is the disparity from the veridical hand to the visual hand.

\subsubsection{Experimental results}

\begin{figure}[htbp]
\centering
\subfigure[Weight changing]{
\includegraphics[width=2in]{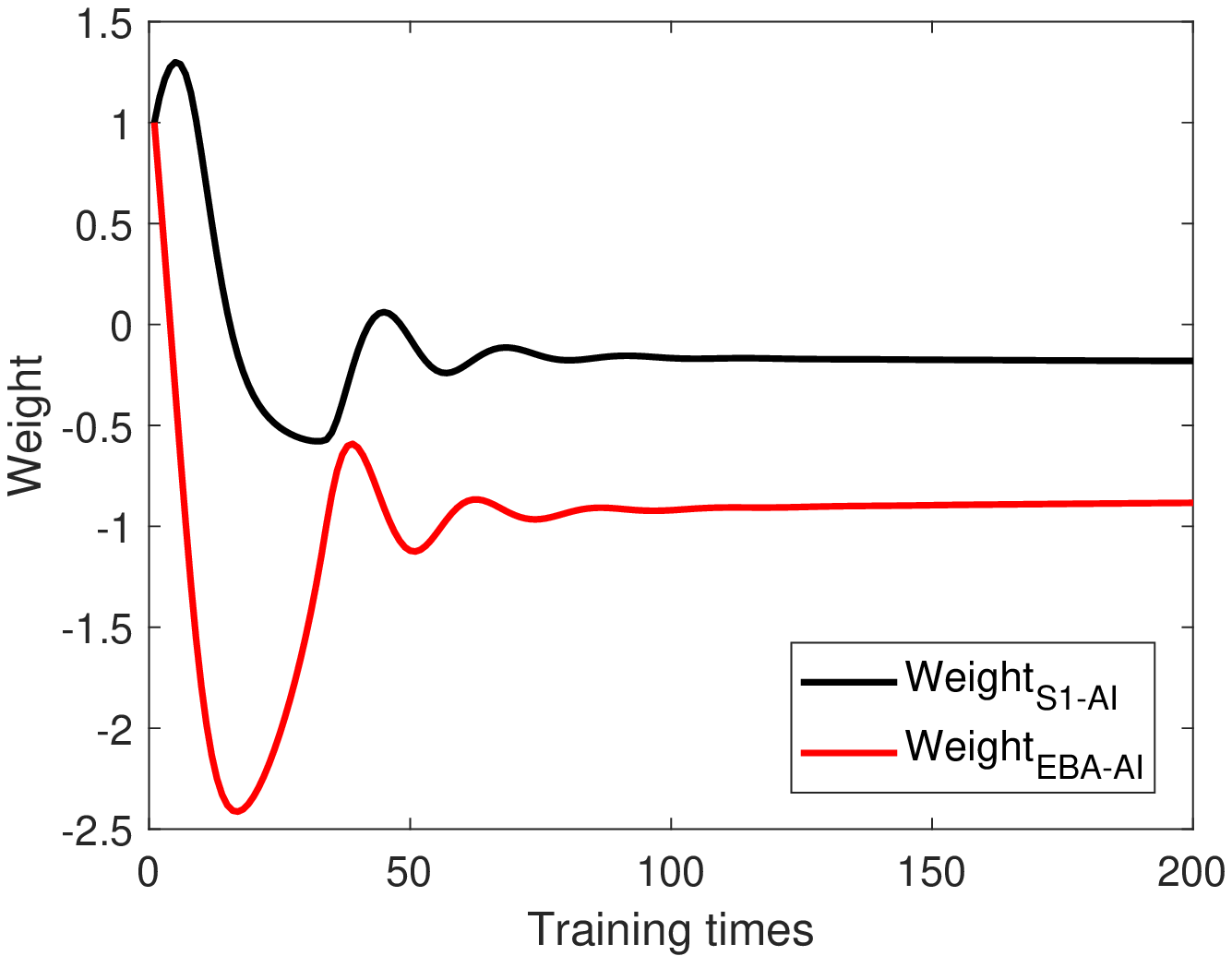}
}
\quad
\subfigure[Behavioral result]{
\includegraphics[width=2in]{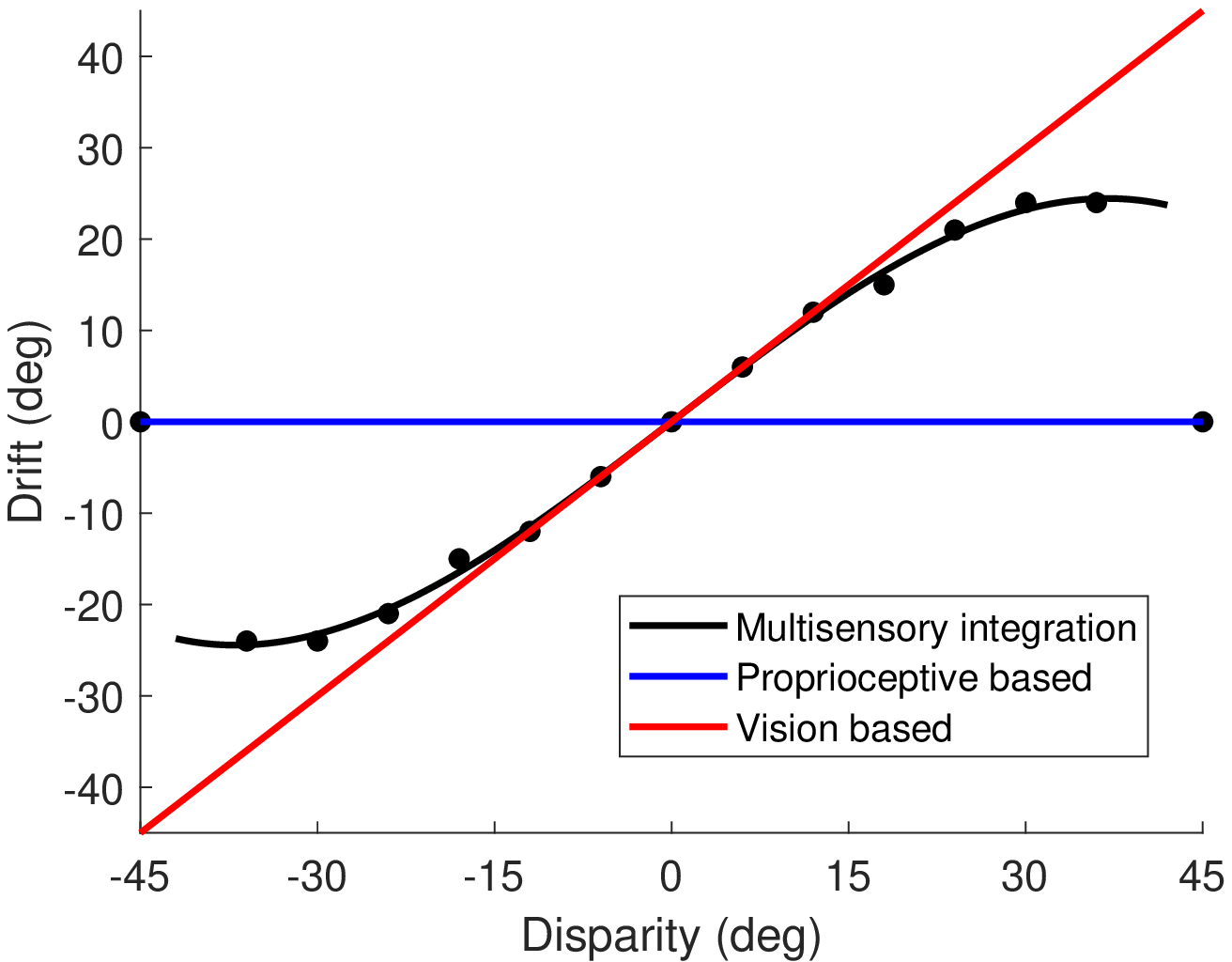}
}
\quad
\subfigure[Small disparity angle]{
\includegraphics[width=1.8in]{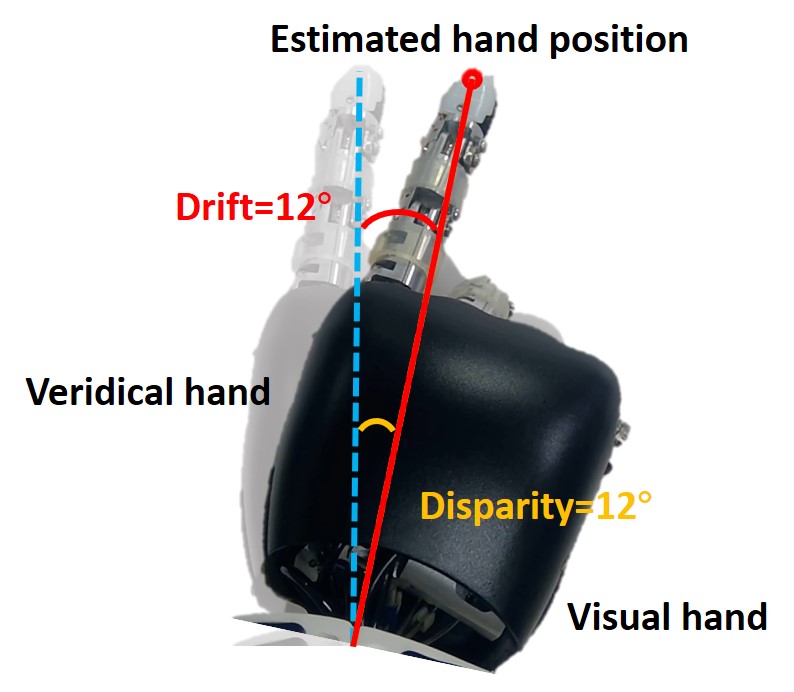}
}
\quad
\subfigure[Medium disparity angle]{
\includegraphics[width=1.8in]{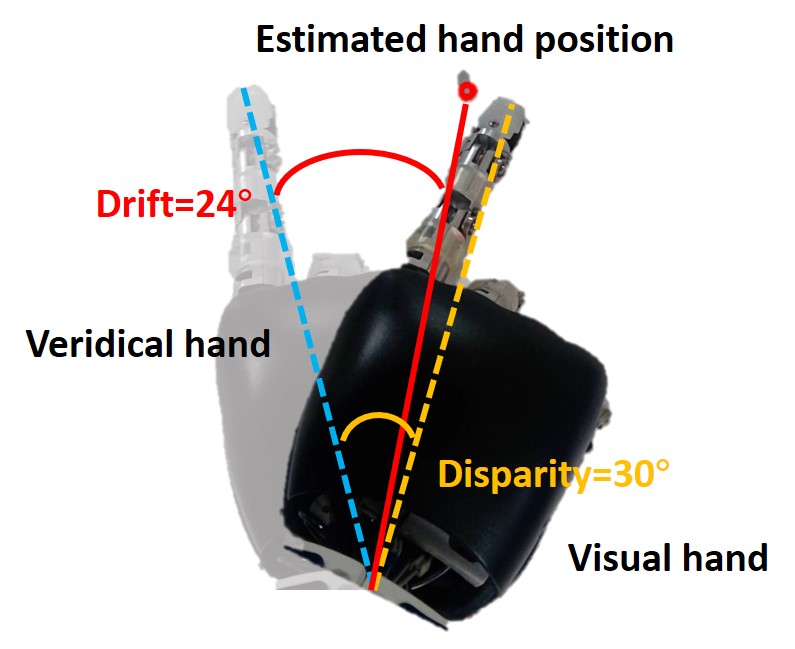}
}
\quad
\subfigure[Large disparity angle]{
\includegraphics[width=1.8in]{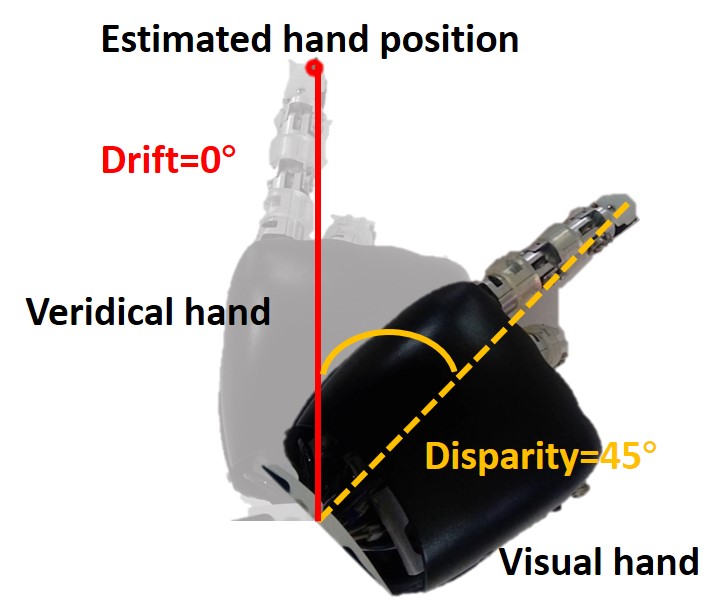}
}
\caption{\textbf{(a)} Weight changing. \textbf{(b)} Behavioral result. The black dots represent the results of behavioral experiments, and the black line represents the curve-fitting results of the behavioral data (Multisensory integration). The red line represents the proprioceptive drift that completely depends on the visual information (Vision based). The blue line represents the proprioceptive drift that completely depends on the proprioceptive information (Proprioceptive based). The horizontal axis is the disparity degree between the visual hand and the veridical hand, and the vertical axis is the degree of proprioceptive drift. \textbf{(c)} Small disparity angle. \textbf{(d)} Medium disparity angle. \textbf{(e)} Large disparity angle.}
\label{RobotW}
\end{figure}

Before the rubber hand illusion experiment, the robot needs to construct the bodily self-perception model through training. During training process, the robot observes its veridical hand directly and trains through random movement. The training process does not require any supervised signal. Figure \ref{RobotW}A shows the synaptic weights between S1 and AI, EBA and AI after training, in which both of them have changed from excitatory connection to inhibitory connection. It also shows that the intensity of inhibitory connection from vision (EBA-AI) is higher than that from proprioception (S1-AI).

Figure \ref{RobotW}B shows the behavioral results of the robot experiment. 
When the hand rotation angle is small (small disparity angle), the proprioceptive drift is small. With the increase of the disparity angle, the proprioceptive drift increases, indicating that the robot mainly relies on visual information for decision-making.
When the hand rotation angle is medium (medium disparity angle), the proprioceptive drift is medium. With the increase of the disparity angle, the proprioceptive drift increases slowly or remains unchanged, indicating that the robot mainly relies on  proprioceptive information for decision-making.
When the hand rotation angle is large (large disparity angle), the proprioceptive drift is zero. With the increase of the disparity angle, the proprioception deviation does not change, indicating that the robot completely relies on the proprioceptive information for decision-making.

Figure \ref{RobotW}C  shows the results of the behavior at a small disparity angle. The disparity angle is $12^{\circ}$, the proprioceptive drift is $12^{\circ}$. 
Figure \ref{RobotW}D  shows the results of the behavior at a medium disparity angle. The disparity angle is $30^{\circ}$, the proprioceptive drift is $24^{\circ}$.
Figure \ref{RobotW}E  shows the results of the behavior at a large disparity angle. The disparity angle is $45^{\circ}$, the proprioceptive drift is $0^{\circ}$.

\subsection{Experiments in simulated environment}

\subsubsection{Proprioceptive drift and Proprioception accuracy experiments}
\begin{figure}[htbp]
\centering
\subfigure[Proprioceptive drift experimental result]{
\includegraphics[width=2.9in]{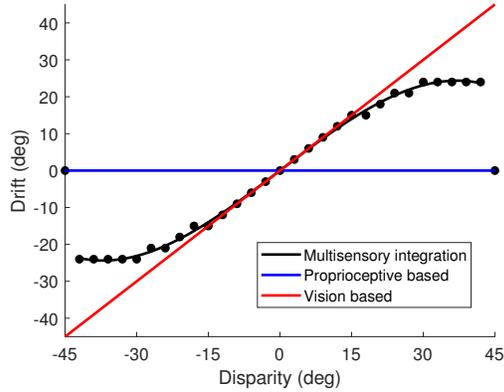}
}
\quad
\subfigure[Proprioception accuracy experimental result]{
\includegraphics[width=2.9in]{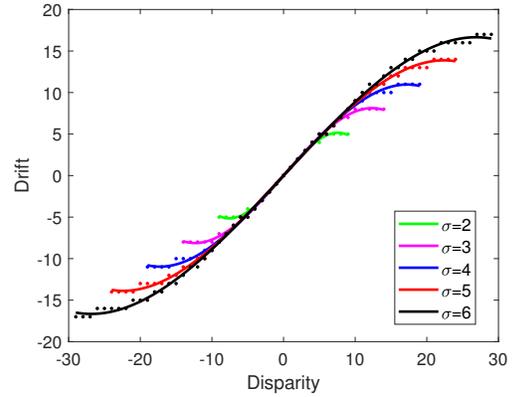}
}
\caption{\textbf{(a)} Proprioceptive drift experiment result. \textbf{(b)} Proprioception accuracy experiment result.}
\label{PD}
\end{figure}

In the simulated environment, the rotation angle ranges from $-60^{\circ}$ to $60^{\circ}$, and the interval is $3^{\circ}$.
The result of the proprioceptive drift experiment is shown in Figure \ref{PD}A, which is similar to the results of the rubber hand illusion behavior experiment in macaques and humans \cite{RN580}. The behavior experiment results in \cite{RN580} show that the proprioceptive drifts increased for small levels of disparity, while it plateaus or even decreases when the disparity exceeds $20^{\circ}$. In our experiment, the proprioceptive drift increases rapidly when the disparity is less than $21^{\circ}$, and became flatter when the disparity is greater than $21^{\circ}$. The results show that when the disparity between the visual hand and the robot's veridical hand is small, the hand position perceived by the robot will be closer to the position of the visual hand, while when the disparity is large, the robot's perception of the hand position is more dependent on the hand position of proprioception.

A recent study has shown that the proprioception accuracy may influence the proneness to the rubber hand illusion, but there are fewer relevant studies \cite{RN978}. In our model, the accuracy of proprioception can be achieved by controlling the standard deviation of the receptive field of the neuron model. The larger the standard deviation, the wider the receptive field range, and the lower the accuracy. On the contrary, the smaller the standard deviation, the narrower the receptive field range, and the higher the accuracy. Therefore, we test the effect of proprioception accuracy on rubber hand illusion by controlling the standard deviation of the receptive field of the neuron, and the result is shown in Figure \ref{PD}B. It shows that when the proprioception accuracy is high (that is, the standard deviation of the receptive field is small), the robot can only produce illusions within a small disparity range, and when the proprioception accuracy is low (that is, the standard deviation of the receptive field is larger), the robot can produce illusions within a large disparity range. That is, the experimental results of the model prove that lower proprioception accuracy is more likely to induce rubber hand illusion, while higher proprioception accuracy is more difficult to induce. This conclusion is consistent with the recent studies \cite{RN978,RN966}. It should be noted that in order to make the experimental results more intuitive, we unified the behavioral results of different proprioceptive accuracies into a scale range.

In the proprioceptive drift experiment, the robot's behavior is influenced by the dynamic change of the neurons' firing rate in the model. In our model, multisensory integration takes place in two areas: TPJ and AI. 
TPJ is the low-level area of multisensory integration, which realizes the initial integration of proprioceptive information and visual information. The synaptic weights of $W_{S1-TPJ}$ and $W_{EBA-TPJ}$ are the same, but since the perception of visual information is caused by movement, TPJ is more affected by visual information during multisensory integration. 
AI is a high-level area of multisensory integration. It integrates the proprioceptive information from S1, the visual information from EBA, and the initial proprioceptive-visual integration information from TPJ to achieve the final integration of multisensory information, and the integration result affects the behavior of the robot. After training, $W_{S1-AI}$ and $W_{EBA-AI}$ change from the initial excitatory connections to inhibitory connections. The inhibitory strength of $W_{EBA-AI}$ is greater than that of $W_{S1-AI}$, which means that AI has a strong inhibitory effect on visual information during multisensory integration (Figure \ref{RobotW}A).

\begin{figure}[htbp]
\centering
\subfigure[Firing rate of the neurons in TPJ]{
\includegraphics[width=2.9in]{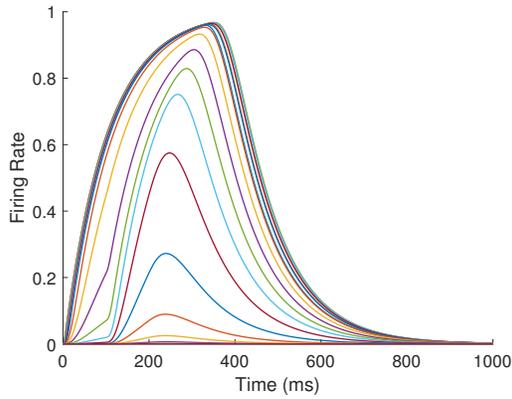}
}
\quad
\subfigure[Firing rate of the neurons in AI]{
\includegraphics[width=2.9in]{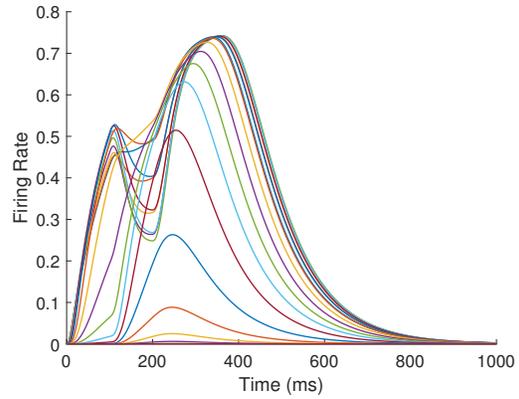}
}
\quad
\subfigure[Local amplification of the neurons in TPJ]{
\includegraphics[width=2.9in]{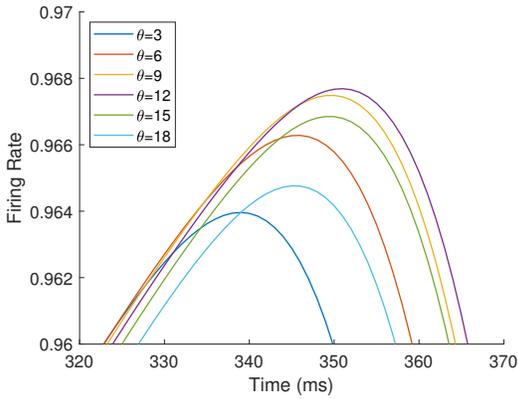}
}
\quad
\subfigure[Local amplification of the neurons in AI]{
\includegraphics[width=2.9in]{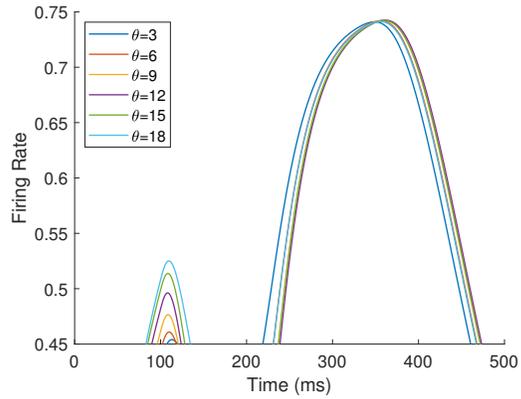}
}
\caption{The firing rate of the neurons in TPJ and AI in the proprioceptive drift experiment (Multisensory integration - Visual dominance). The proprioceptive perception is $0^{\circ}$, the visual disparity is $12^{\circ}$. The proprioception drift is $12^{\circ}$ after multisensory integration.
\textbf{(a and b)} The firing rate of the neurons in TPJ and AI. \textbf{(c and d)} The results of local amplification near the maximum peak firing rates of the neurons in TPJ and AI. Here the six neurons with the highest firing rates are selected and the dynamic changes of these neurons are plotted.}
\label{NM}
\end{figure}

The firing rate of the TPJ and AI in the proprioceptive drift experiment (Multisensory integration - Visual dominance) is shown in Figure \ref{NM}.
When the visual disparity is small, the receptive field overlap of proprioceptive information and visual information is huge, so the TPJ area presents a single peak, visual-information-dominated multisensory integration result. The results of multisensory integration when the proprioceptive perception is $0^{\circ}$, and the visual disparity is $12^{\circ}$ are shown in Figure \ref{NM}A and Figure \ref{NM}C.
Figure \ref{NM}C shows the dynamic changes of the six neurons with the highest firing rate in TPJ. The firing rate from high to low is $12^{\circ}, 9^{\circ}, 15^{\circ}, 6^{\circ}, 18^{\circ}, 3^{\circ}$, and the initial integration result in TPJ is $12^{\circ}$.
The AI area presents multisensory integration results of bimodal peaks (as shown in Figure \ref{NM}B and Figure \ref{NM}D). 
The anterior peak is mainly inhibited by neurons near the proprioceptive perception $0^{\circ}$ and the visual perception $12^{\circ}$. The firing rate from high to low is $18^{\circ}, 15^{\circ}, 12^{\circ}, 9^{\circ}, 6^{\circ}, 3^{\circ}$, and the result of anterior peak integration is $18^{\circ}$.
In the posterior peak, the inhibition intensity decreases with the disappearance of the proprioceptive and visual stimulation, and the firing rate is $12^{\circ}, 9^{\circ}, 15^{\circ}, 6^{\circ}, 18^{\circ}, 3^{\circ}$ from high to low. The firing rate of neurons in the posterior peak is higher than that in the anterior peak, and the final integration result in AI is $12^{\circ}$.

\begin{figure}[htbp]
\centering
\subfigure[Firing rate of the neurons in TPJ]{
\includegraphics[width=2.9in]{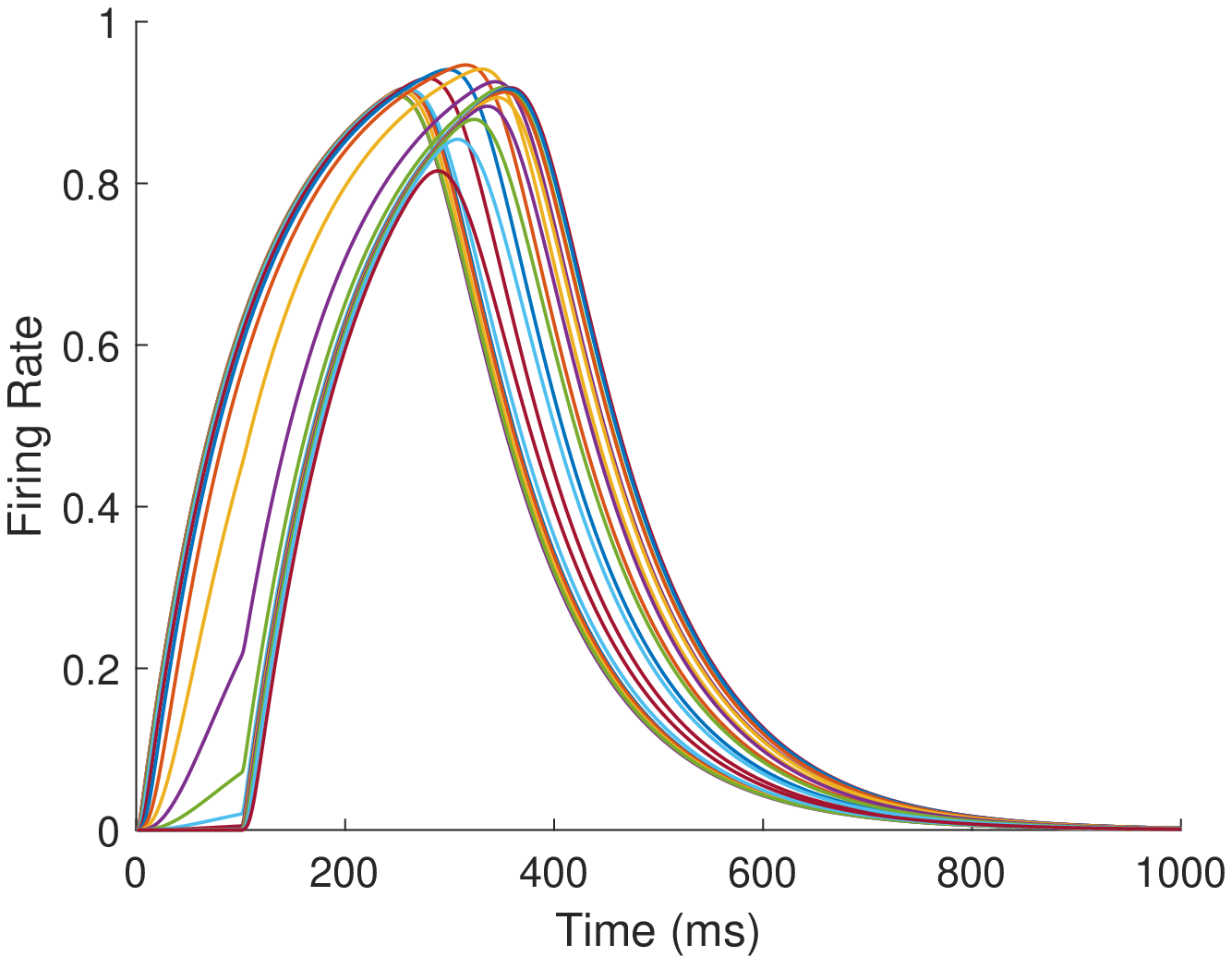}
}
\quad
\subfigure[Firing rate of the neurons in AI]{
\includegraphics[width=2.9in]{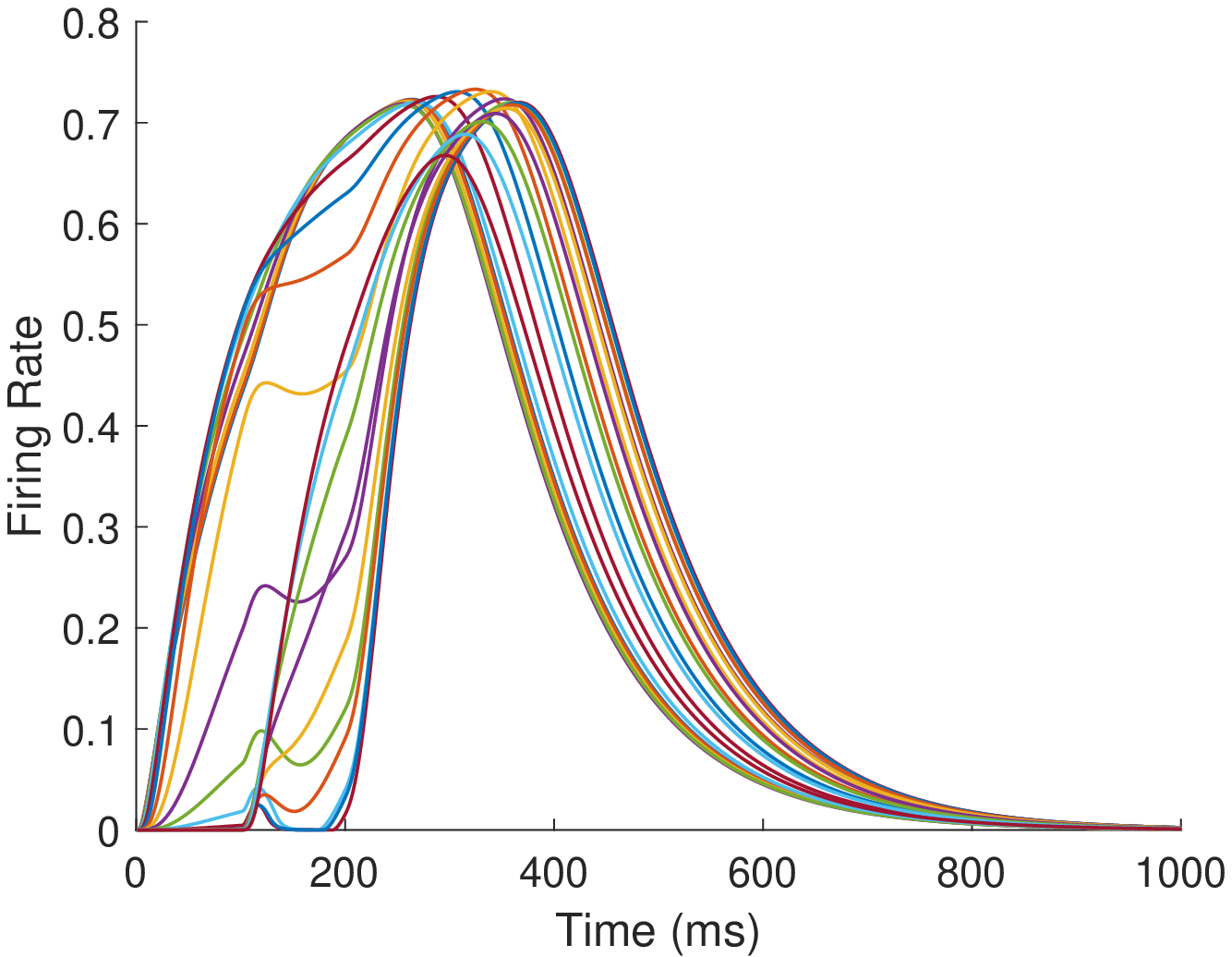}
}
\quad
\subfigure[Local amplification of the neurons in TPJ]{
\includegraphics[width=2.9in]{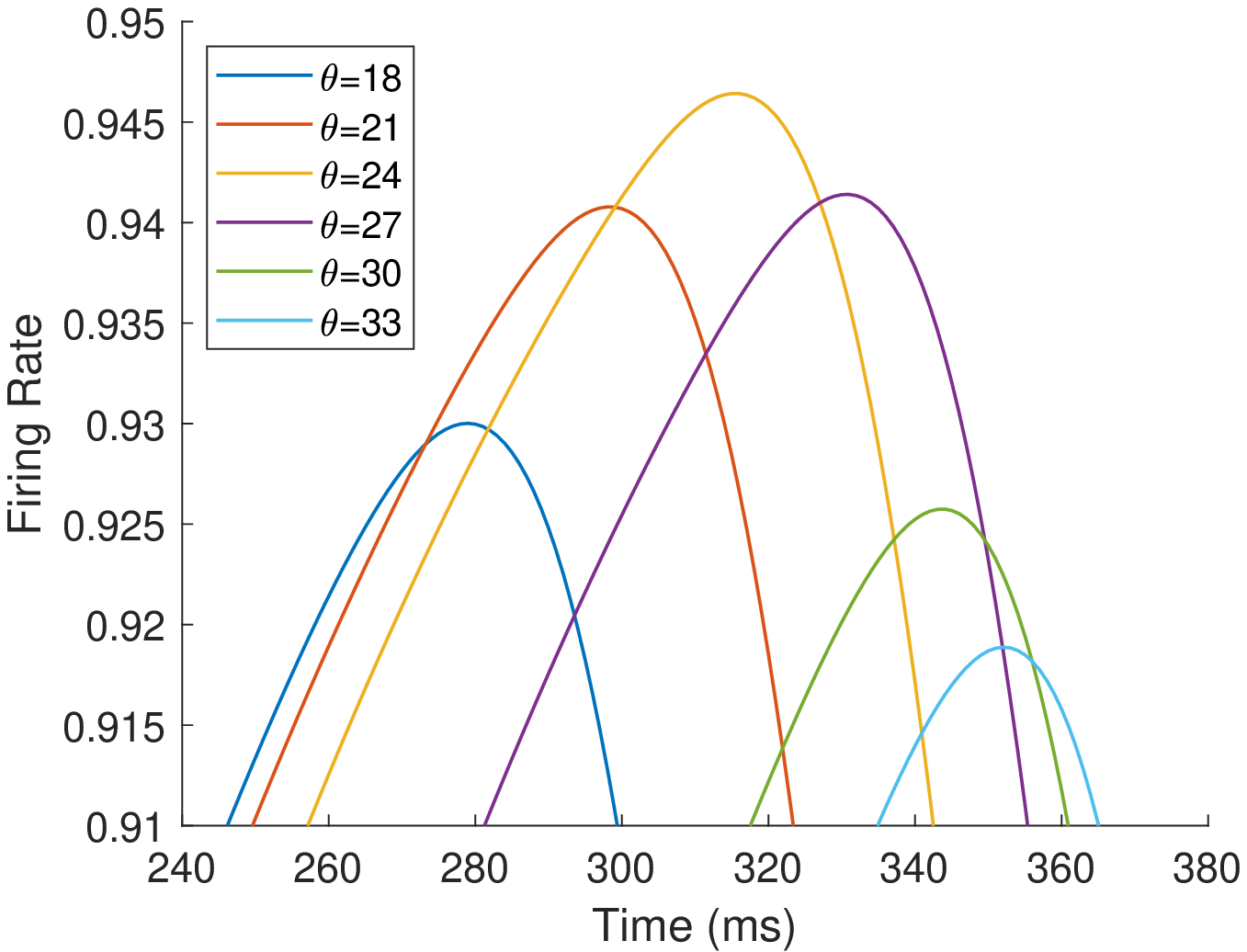}
}
\quad
\subfigure[Local amplification of the neurons in AI]{
\includegraphics[width=2.9in]{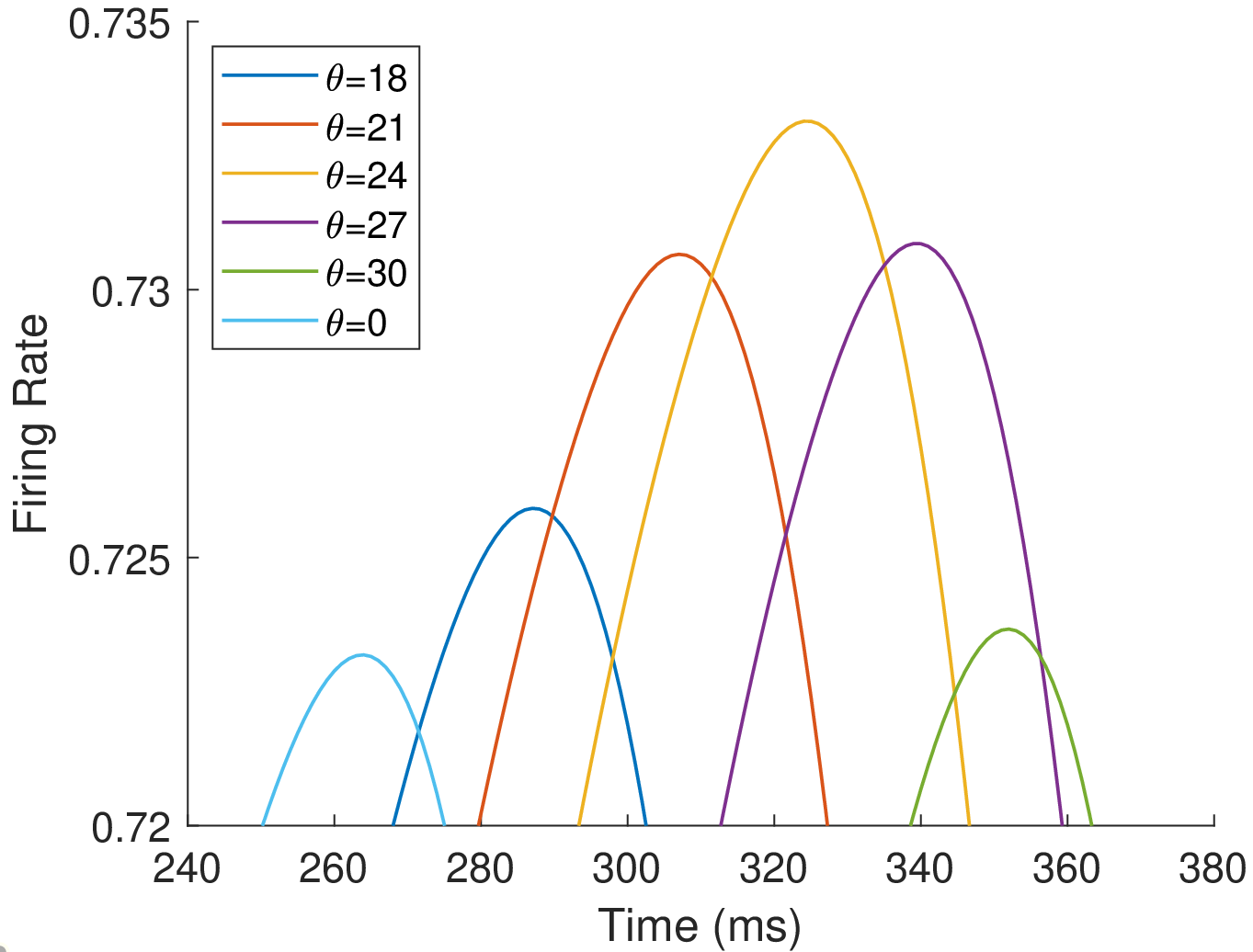}
}
\caption{The firing rate of the neurons in TPJ and AI in the proprioceptive drift experiment (Multisensory integration - Proprioception dominance). The proprioceptive perception is $0^{\circ}$, the visual disparity is $39^{\circ}$. The proprioception drift is $24^{\circ}$ after multisensory integration.
\textbf{(a and b)} The firing rate of the neurons in TPJ and AI. \textbf{(c and d)} The results of local amplification near the maximum peak firing rates of the neurons in TPJ and AI. Here the six neurons with the highest firing rates are selected and the dynamic changes of these neurons are plotted.}
\label{NMP}
\end{figure}

The firing rate of the TPJ and AI in the proprioceptive drift experiment (Multisensory integration - Proprioception dominance) is shown in Figure \ref{NMP}. When the visual disparity is moderate, the information integration in TPJ mainly occurs at the location where the proprioceptive information and visual information receptive fields overlap. The TPJ area presents multi-peak, proprioceptive-information-dominated multisensory integration result. The results of multisensory integration when the proprioceptive perception is $0^{\circ}$, and the visual disparity is $39^{\circ}$ are shown in Figure \ref{NMP}A and Figure \ref{NMP}C.
Figure \ref{NMP}C shows the dynamic changes of the six neurons with the highest firing rate in TPJ. The firing rate from high to low is $24^{\circ}, 27^{\circ}, 21^{\circ}, 18^{\circ}, 30^{\circ}, 33^{\circ}$, and the initial integration result in TPJ is $24^{\circ}$.
The AI area presents multi-peak multisensory integration results (as shown in Figure \ref{NMP}B and Figure \ref{NMP}D). 
Figure \ref{NMP}B shows the dynamic changes of all neurons.
Figure \ref{NMP}D shows the dynamic changes of the six neurons with the highest firing rate in AI. The firing rate from high to low is $24^{\circ}, 27^{\circ}, 21^{\circ}, 18^{\circ}, 30^{\circ}, 0^{\circ}$. The final integration result in AI is $24^{\circ}$.

\begin{figure}[htbp]
\centering
\subfigure[Firing rate of the neurons in TPJ]{
\includegraphics[width=2.9in]{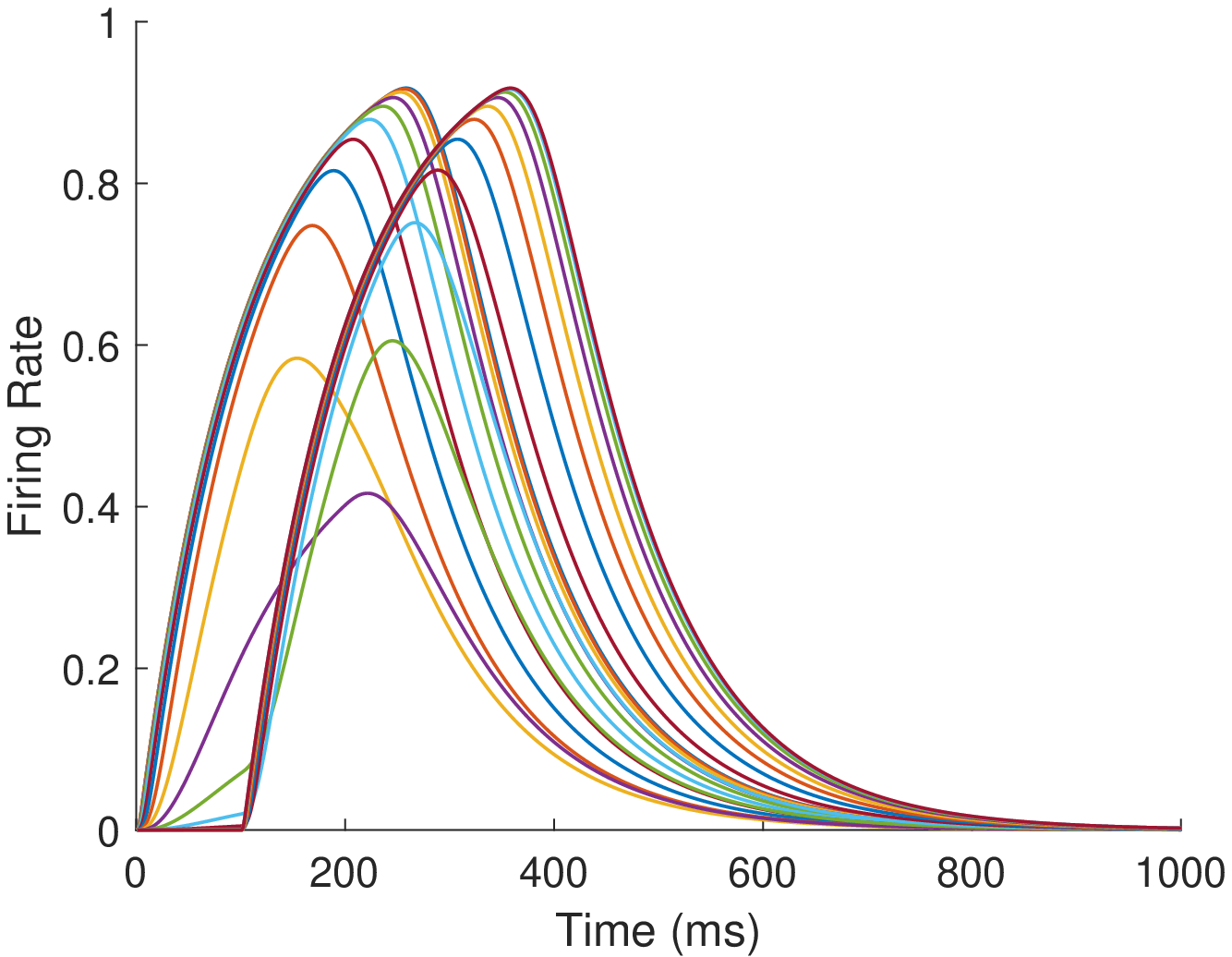}
}
\quad
\subfigure[Firing rate of the neurons in AI]{
\includegraphics[width=2.9in]{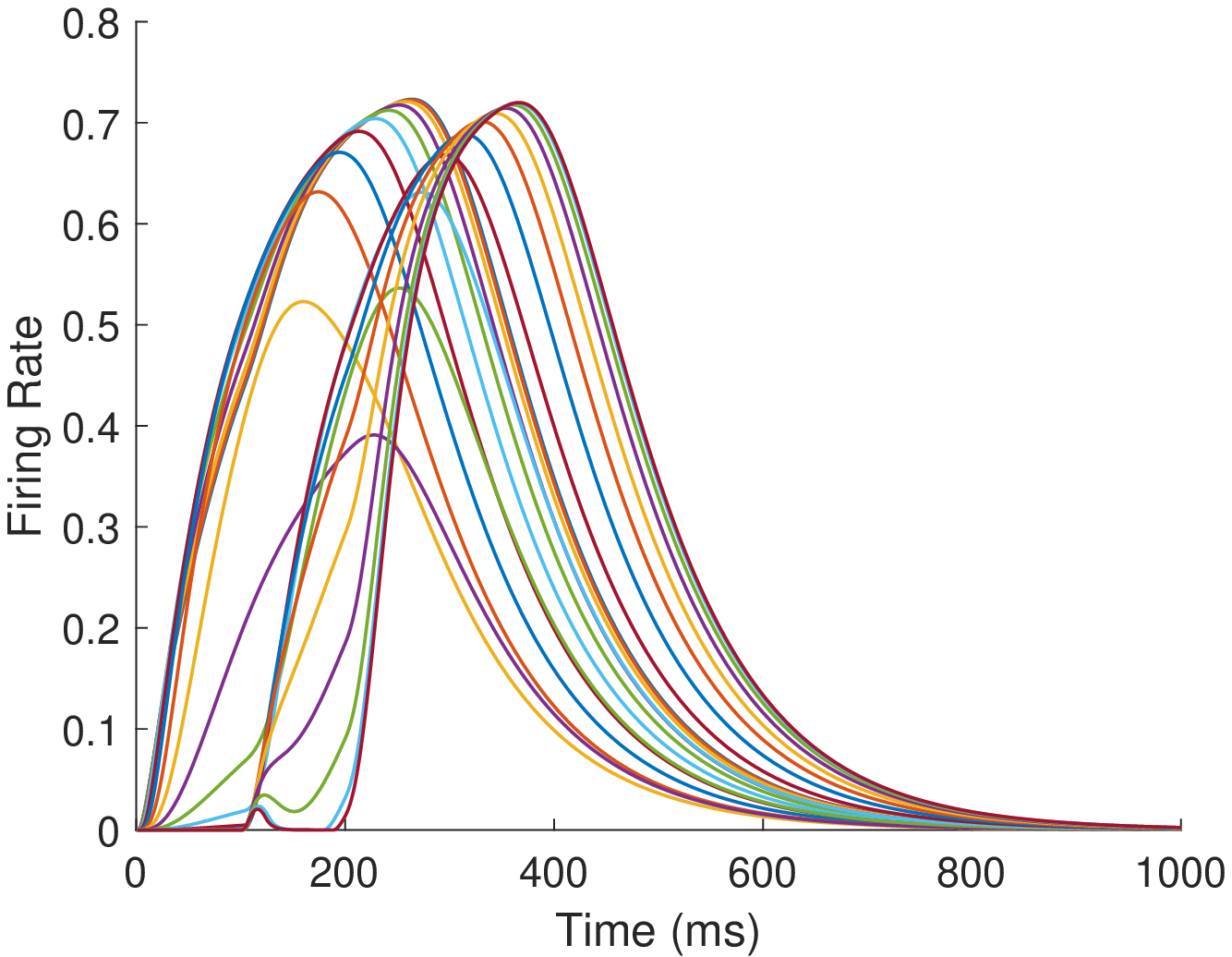}
}
\quad
\subfigure[Local amplification of the neurons in TPJ]{
\includegraphics[width=2.9in]{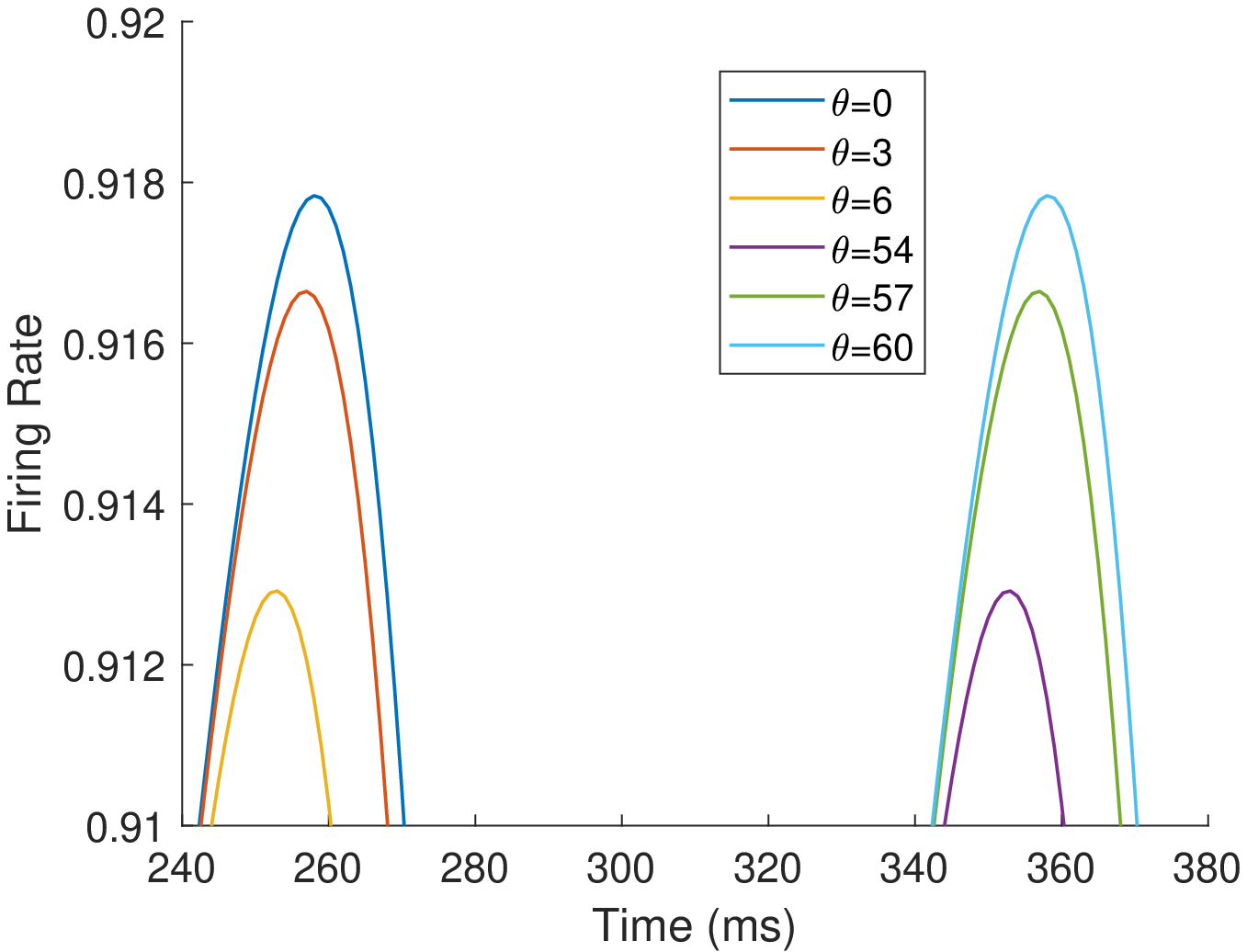}
}
\quad
\subfigure[Local amplification of the neurons in AI]{
\includegraphics[width=2.9in]{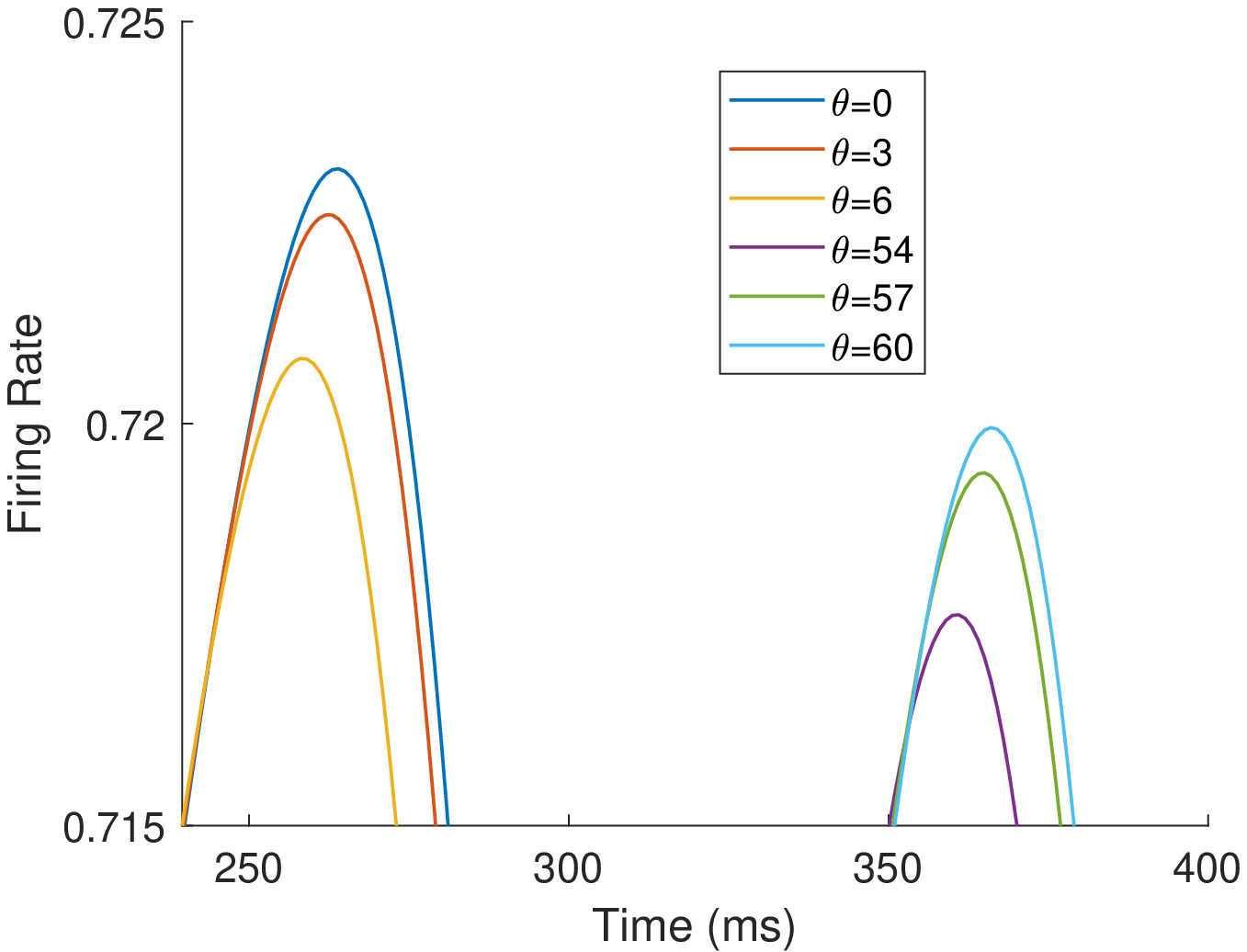}
}
\caption{The firing rate of the TPJ and AI in the proprioceptive drift experiment (Multisensory integration - Proprioception based). The proprioceptive perception is $0^{\circ}$, and the visual disparity is $60^{\circ}$. The proprioception drift is $0^{\circ}$ after multisensory integration.
\textbf{(a and b)} The firing rate of the neurons in TPJ and AI. \textbf{(c and d)} The results of local amplification near the maximum peak firing rates of the neurons in TPJ and AI. Here the six neurons with the highest firing rates are selected and the dynamic changes of these neurons are plotted. }
\label{NP}
\end{figure}

The firing rate of the TPJ and AI in the proprioceptive drift experiment (Multisensory integration - Proprioception based) is shown in Figure \ref{NP}.
When the visual disparity is large, the overlapping area of proprioceptive neurons and visual neurons' receptive fields is small, and the integration of TPJ is weak. Therefore, the TPJ presents completely doublet peak results, with the anterior peak dominated by proprioceptive information and the posterior peak dominated by visual information.
The results of multisensory integration when the proprioceptive perception is $0^{\circ}$, and the visual disparity is $60^{\circ}$ are shown in Figure \ref{NP}A and Figure \ref{NP}C.
Figure \ref{NP}C shows the dynamic changes of the six neurons with the highest firing rate in TPJ. It shows that the anterior peak presents the integration result dependent on proprioception information, and the firing rate from high to low is $0^{\circ}, 3^{\circ}, 6^{\circ}$. The posterior peak presents the integration result dependent on visual information, and the firing rate from high to low is $60^{\circ}, 57^{\circ}, 54^{\circ}$. The firing rates of the six neurons are $60^{\circ}, 0^{\circ}, 57^{\circ}, 3^{\circ}, 54^{\circ}, 6^{\circ}$ from high to low, and the initial integration result in TPJ is $60^{\circ}$.
The AI area presents doublet peak multisensory integration results (as shown in Figure \ref{NP}B and Figure \ref{NP}D). 
Since the inhibition weight of visual connections ($W_{EBA-AI}$) in the AI area is greater than that of proprioception ($W_{S1-AI}$), the anterior peak that relies on proprioception is used as the final output.
The firing rate from high to low is $0^{\circ}, 3^{\circ}, 6^{\circ}, 60^{\circ}, 57^{\circ}, 54^{\circ}$. The final integration result in AI is $0^{\circ}$.

\subsubsection{Appearance replacement and Asynchronous experiments}
\begin{figure}[htbp]
\centering
\subfigure[Appearance replacement experiment result]{
\includegraphics[width=2.9in]{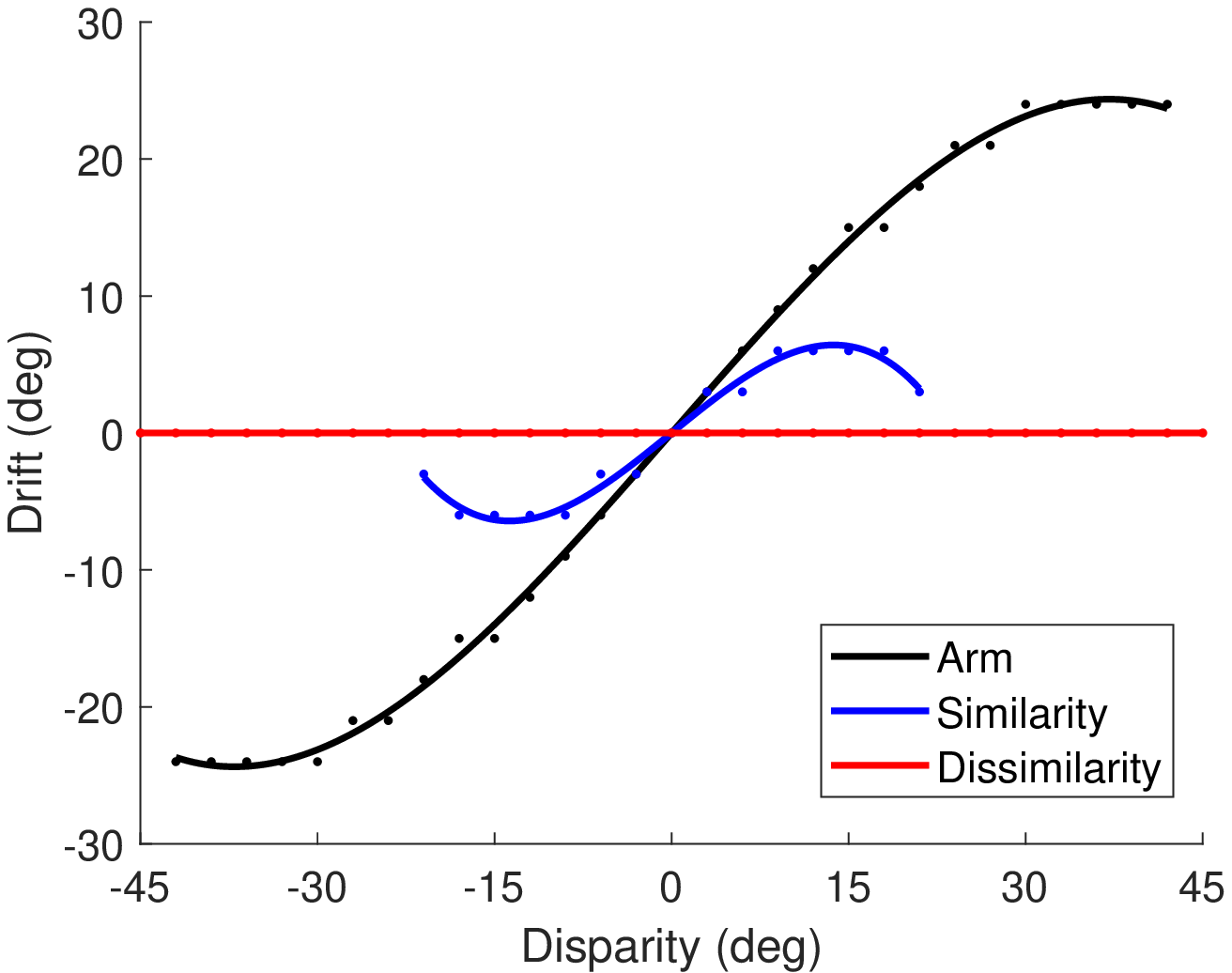}
}
\quad
\subfigure[Asynchronous experiment result]{
\includegraphics[width=2.9in]{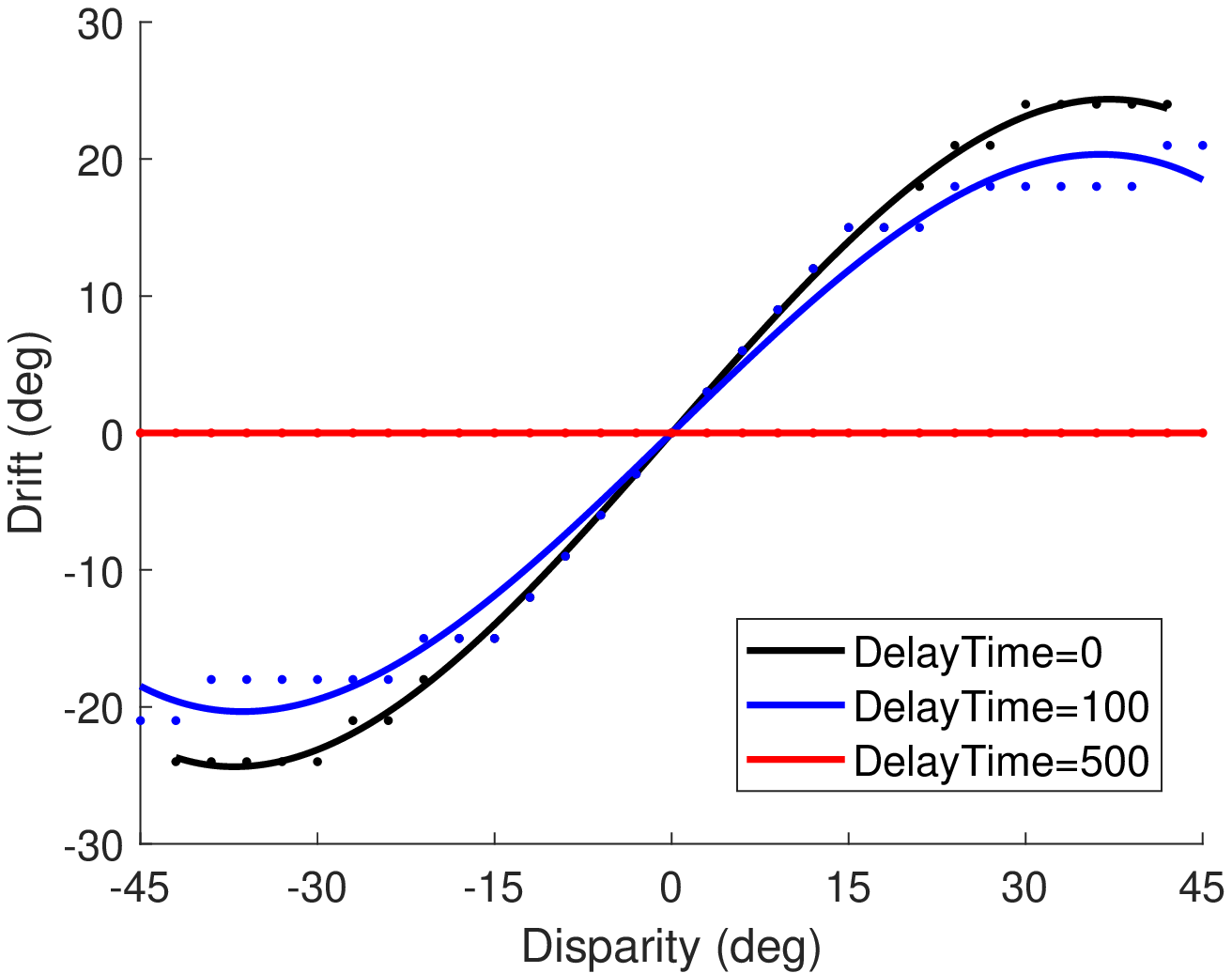}
}
\caption{The results of appearance replacement experiment and asynchronous experiment.}
\label{AD}
\end{figure}

The result of the appearance replacement experiment is shown in Figure \ref{AD}A, which is similar to the results of the prior knowledge of body representation behavior experiment (that is, replacing the visual hand in the rubber hand illusion with a wooden block) in macaques and humans \cite{RN580}. The experimental result in \cite{RN580} shows that the proprioceptive drifts in the wood condition is significantly reduced than that in the visual hand condition. In our experiment, 
in the case of high similarity (that is, the hand in the robot's field of vision is similar to its own hand), the proprioceptive drift of the robot is significantly smaller than that of the own visual hand. In addition, the range of visual disparity that can induce the robot's rubber hand illusion is narrower. When the disparity exceeded $21^{\circ}$, the robot's proprioceptive drift is $0^{\circ}$, and the rubber hand illusion disappears.
In the case of dissimilarity (that is, the hand in the robot's field of vision is completely different from its own hand), the robot's proprioceptive drift is $0^{\circ}$, indicating that the robot does not consider the hand in the field of view as its own. At the neuronal scale, the response of neurons in EBA area to a similar hand is lower than that of their own hand, which leads to the weak contribution of visual information for multisensory integration, with TPJ and AI relying more on proprioceptive information for integration; the neurons in the EBA area respond much less to dissimilar hands than their own hand, further weakening the contribution of visual information for multisensory integration in the TPJ and AI areas.

The studies in \cite{RN980,RN979} show that the majority of participants would not experience the rubber hand illusion if the asynchrony is greater than 500 or 300 ms. Here we test the effect of synchrony on the rubber hand illusion. The result of the asynchronous experiment is shown in Figure \ref{AD}B. When the delay time is 100 ms, the proprioceptive drift of the robot is smaller than that of the own visual hand. When the delay time is 500 ms, the robot's proprioceptive drift is $0^{\circ}$, indicating that the robot does not think that the hand in the field of view is its own. At the neuronal scale, the delayed presentation of visual information will weaken the contribution of visual information to multisensory integration, and a large delay will result in the inability to integrate proprioceptive information and visual information within the effective firing time window of neurons.

\subsubsection{Proprioception only and Vision only experiments}
When there is only proprioception information, that is, when the robot only moves its own hand and no hand is observed in the field of vision, the AI neuron is fired, indicating that the robot thinks that the moving hand belongs to itself. When there is only visual information, that is, the robot's own hand does not move, but the hand movement can be seen in the field of vision, the neuron in AI is not fired, indicating that the robot does not think that the moving hand belongs to itself. At the neuronal scale, the main reason for this result is that the inhibition intensity of visual information in AI is stronger than that of proprioceptive information.

\subsubsection{Disability experiment}

We explore the effects of TPJ disability and AI disability on the rubber hand illusion by setting synaptic weights between different brain areas. Take the Proprioception accuracy experiment as an example.

\begin{figure}[htbp]
\centering
\subfigure[TPJ disability]{
\includegraphics[width=2.9in]{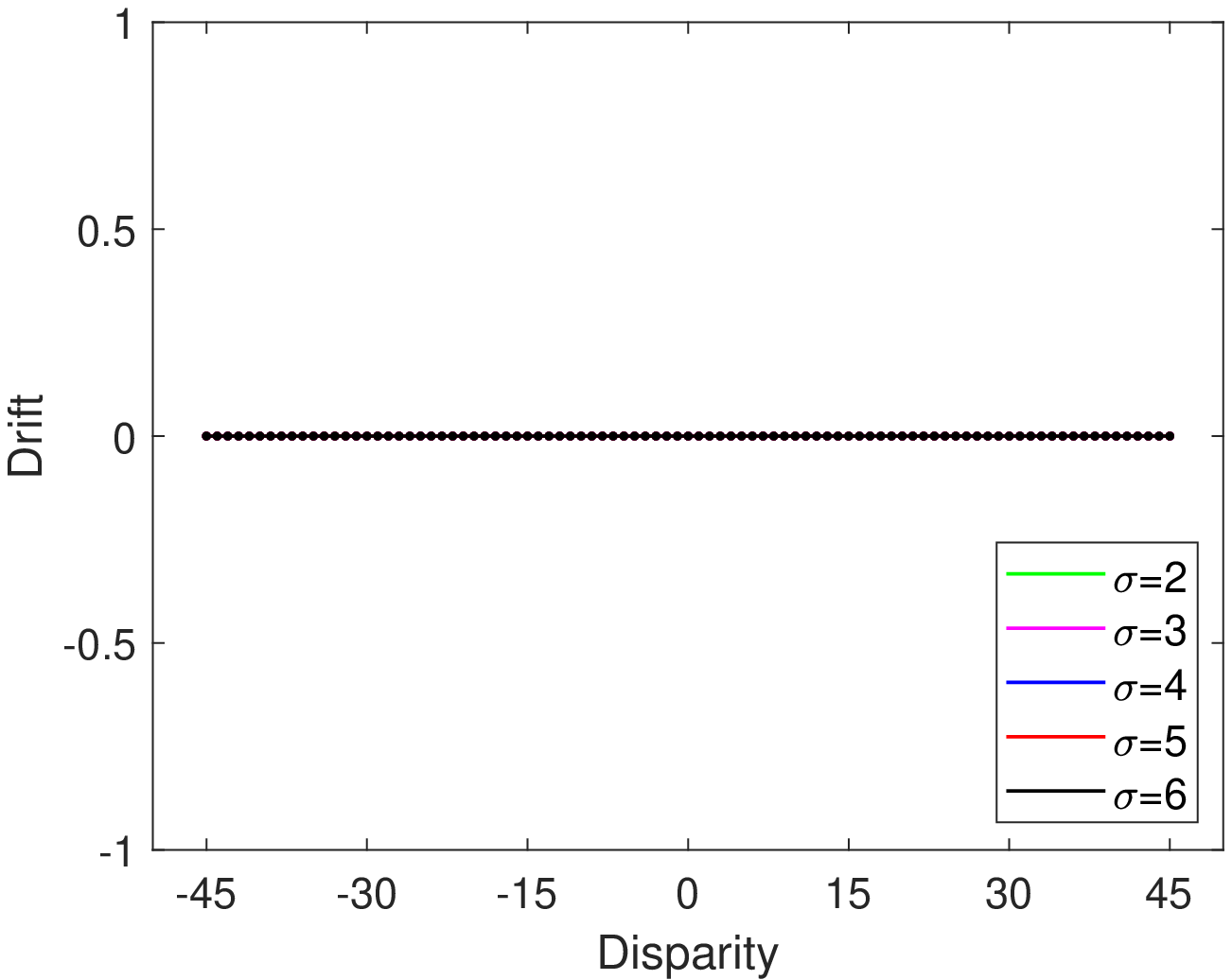}
}
\quad
\subfigure[AI disability]{
\includegraphics[width=2.9in]{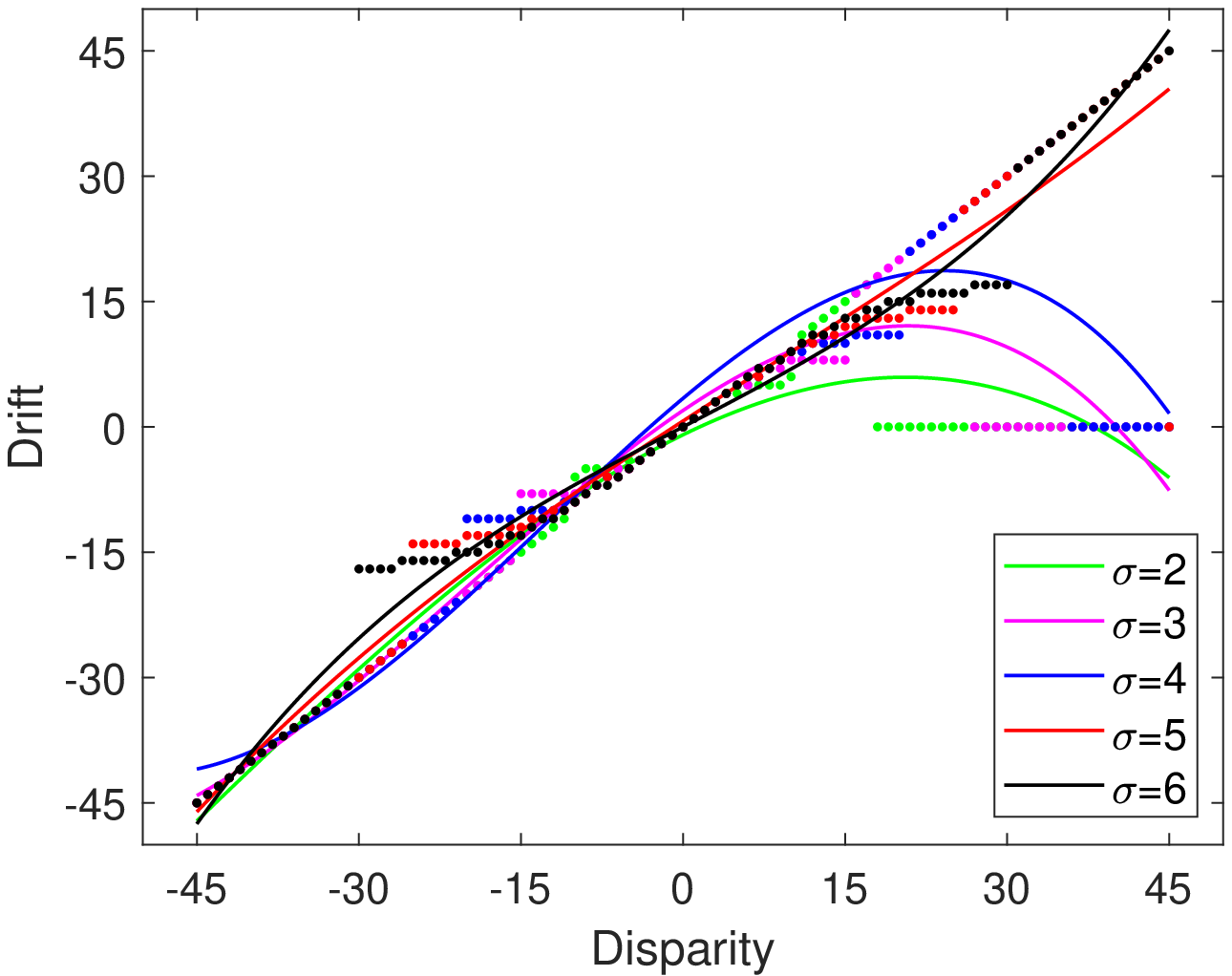}
}
\caption{The result of disability experiment}
\label{Dis}
\end{figure}

In the TPJ disability experiment, we set $W_{S1-TPJ}$ and $W_{EBA-TPJ}$ as 0. Multisensory information is integrated by the AI only, and the TPJ no longer integrated any information. After training and testing, the experimental result of TPJ disability with different proprioception accuracy is shown in Figure \ref{Dis}A. This result indicates that when TPJ is completely disabled, multisensory information integration by AI alone cannot induce the rubber hand illusion. In addition, we simulate the impact of varying extents of TPJ disability on the rubber hand illusion by setting different synaptic weights. The experimental results show that when the extent of TPJ disability is low (e.g., the values of $W_{S1-TPJ}$ and $W_{EBA-TPJ}$ are large), TPJ is still able to integrate multisensory information, and the robot can be induced rubber hand illusion. But the proprioceptive drift is smaller, and the range of visual disparity that can induce the robot's rubber hand illusion is narrower. As the extent of disability increases (i.e. the values of $W_{S1-TPJ}$ and $W_{EBA-TPJ}$ decrease), the rubber hand illusion becomes more difficult to be induced. The experimental results are consistent with the behavioral experiment results that use the Transcranial Magnetic Stimulation over TPJ to reduce the extent of rubber hand illusion \cite{tsakiris2010my}.

In the AI disability experiment, we set $W_{S1-AI}$, $W_{TPJ-AI}$ and $W_{EBA-AI}$ as 0. Multisensory information is integrated by the TPJ, and the AI no longer integrated any information. After training and testing, the experimental result of AI disability with different proprioception accuracy is shown in Figure \ref{Dis}B. As seen in Figure \ref{Dis}B, although the robot is induced with the rubber hand illusion at partial visual deflection angles, the AI disability robot mainly relies on vision for decision-making at most visual deflection angles. That is, AI disability cannot effectively induce the rubber hand illusion.

The results of the disability experiment indicate that the generation of rubber hand illusion is the result of the joint action of primary multisensory integration area (e.g., TPJ) and high-level multisensory integration area (e.g., AI), and neither is indispensable.

\subsection{Comparison with other models}

Computational models of self-body representation, especially those that can reproduce and explain the rubber hand illusion, are less studied. Researchers have proposed theoretical models of body representation mainly from the perspectives of Predictive coding and Bayesian causal inference \cite{RN982}. Table \ref{mylabel} shows the comparison results with other models.

The core idea of the Predictive coding approaches \cite{RN983,RN987} is to refine the body estimation through the minimization of the errors between perception and prediction. The Active inference models \cite{RN990,RN986} can be regarded as an extension of Predictive coding approaches. These methods can well reproduce and explain the proprioceptive drift experiment, and have been verified in human, robot and simulated environment. However, these studies did not involve various experiments of the rubber hand illusion, nor could they explain the specific computational mechanism of individual neurons and population neurons. 

Bayesian causal inference model is extensively used in theoretical modeling of multimodal integration, and has been repeatedly verified at the behavioral and neuronal levels. The Bayesian causal inference model can well reproduce and explain a variety of rubber hand illusion experiments\cite{RN970,RN580}. However, in the experiment of rubber hand illusion, most of the Bayesian causal inference models have problems similar to the Predictive coding approaches, such as the explanatory scale remains at the behavioral scale, and also does not explain how the rubber hand illusion is generated from the neuron scale. In addition, there are also some studies on neural network modeling, although they mainly focus on multimodal integration and do not involve rubber hand illusion experiments \cite{RN991}. 

Compared with these models, we build a Brain-inspired bodily self-perception model from the perspective of brain-inspired computing. Our model can not only reproduce as many as six experiments of the rubber hand illusion, but also reasonably explain the rubber hand illusion from the neuron scale at the same time, which is helpful to reveal the computational and neural mechanism of the rubber hand illusion. And the computational model of Disability experiment demonstrates that the rubber hand illusion cannot be induced without TPJ, which performs primary multisensory integration, and AI, which performs high-level multisensory integration.

\begin{table}[]
\centering
\caption{Comparison with other models. $O$ represents that the model could reproduce the correlation feature, and $-$ represents that it is not mentioned.}
\renewcommand{\arraystretch}{1.2} 
\label{mylabel}
\begin{threeparttable}
\begin{tabular}{|c|c|c|c|c|c|c|c|}
\hline
Experiment              & \cite{RN983}    & \cite{RN987}  & \cite{RN990}       & \cite{RN986}  & \cite{RN970}   & \cite{RN580}    & Our model  \\ \hline
Proprioceptive drift    & O    & O  & O       & O  & O   & O    & O          \\ \hline
Proprioception accuracy & -    & -  & -       & -  & -   & -    & O          \\ \hline
Appearance replacement  & -    & -  & -       & -  & -   & O    & O          \\ \hline
Asynchronous            & -    & -  & O       & -  & O   & -    & O          \\ \hline
Proprioception only     & -    & -  & -       & -  & O   & -    & O          \\ \hline
Vision only             & -    & -  & -       & -  & O   & -    & O          \\ \hline
Disability  & -    & -  & -       & -  & -   & -    & O          \\ \hline
Model                   & PC   & PC & Deep AIF & AIF & BCI & BCI  & Brain-Self \\ \hline
Participant             & H, R & R  & S       & H  & H   & M, H & R, S          \\ \hline
\end{tabular}
\begin{tablenotes}
        \footnotesize
        \item Model: PC, Predictive coding; AIF, Active inference model; BCI, Bayesian causal inference model; Brain-Self, Brain-inspired Bodily Self-perception Model.
        \item Participant: H, Human; R, Robot; S, simulated environment; M, Monkey.
      \end{tablenotes}
\end{threeparttable}
\end{table}

\section{Discussion}
In this study, we integrate the biological findings of bodily self-consciousness and construct a Brain-inspired bodily self-perception model, which could construct the bodily self-perception autonomously without any supervision signals. It can reproduce six rubber hand illusion experiments, and reasonably explain the causes and results of rubber hand illusion from the neuron scale. Compared with other models, our model explains the computational mechanism that the brain encodes bodily self-consciousness and how the body illusion we subjectively perceive is generated by neural networks. Especially, the experimental results of this model can well fit the behavioral and neural data of monkeys in biological experiments. This model is helpful to reveal the computational and neural mechanism of the rubber hand illusion.

In the biological experiment of rubber hand, Fang et al. \cite{RN580} recorded the firing rate of neurons in the premotor cortex of 2 monkeys in the behavior task. They defined two types of neurons: integration neurons and segregation neurons. If a neuron's firing rate under the visual-proprioceptive congruent task is greater than that under the proprioception-only task, it indicates that the neuron prefers to integrate visual and proprioception information, which they define as `integration neuron'; Conversely, if a neuron's firing rate under proprioception-only task is greater than that under visual-proprioceptive congruent task, it means that the neuron prefers only proprioception information, which they define as `segregation neuron'. By analyzing the firing rates of neurons under different tasks, they found that the firing rates of integration neurons decreased with increasing visual disparity during the target-holding period, and the firing rates of segregation neurons increased with increasing visual disparity during the preparation period.

\begin{figure}[htbp]
\centering
\subfigure[Firing rate of the TPJ]{
\includegraphics[width=2.9in]{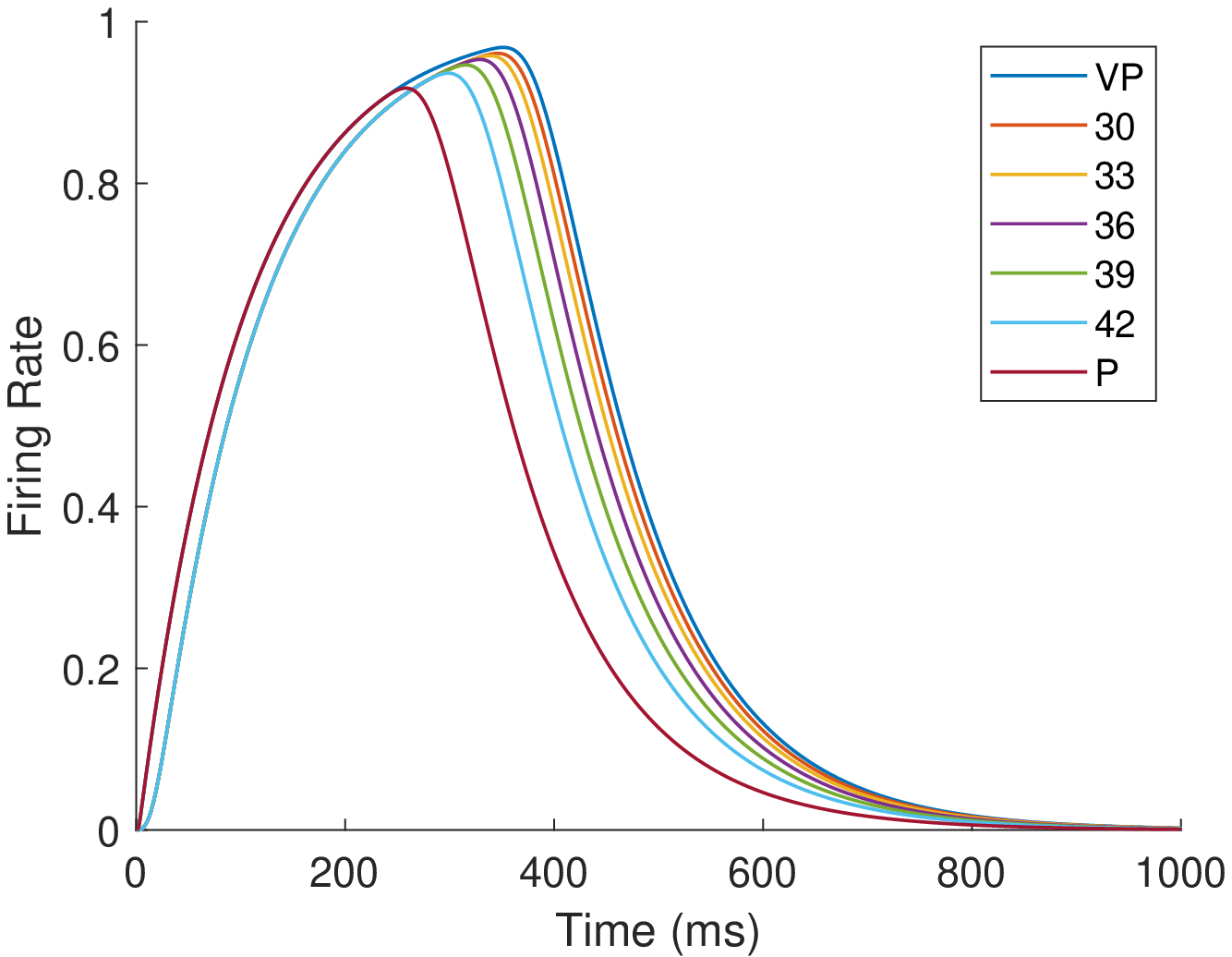}
}
\quad
\subfigure[Firing rate of the AI]{
\includegraphics[width=2.9in]{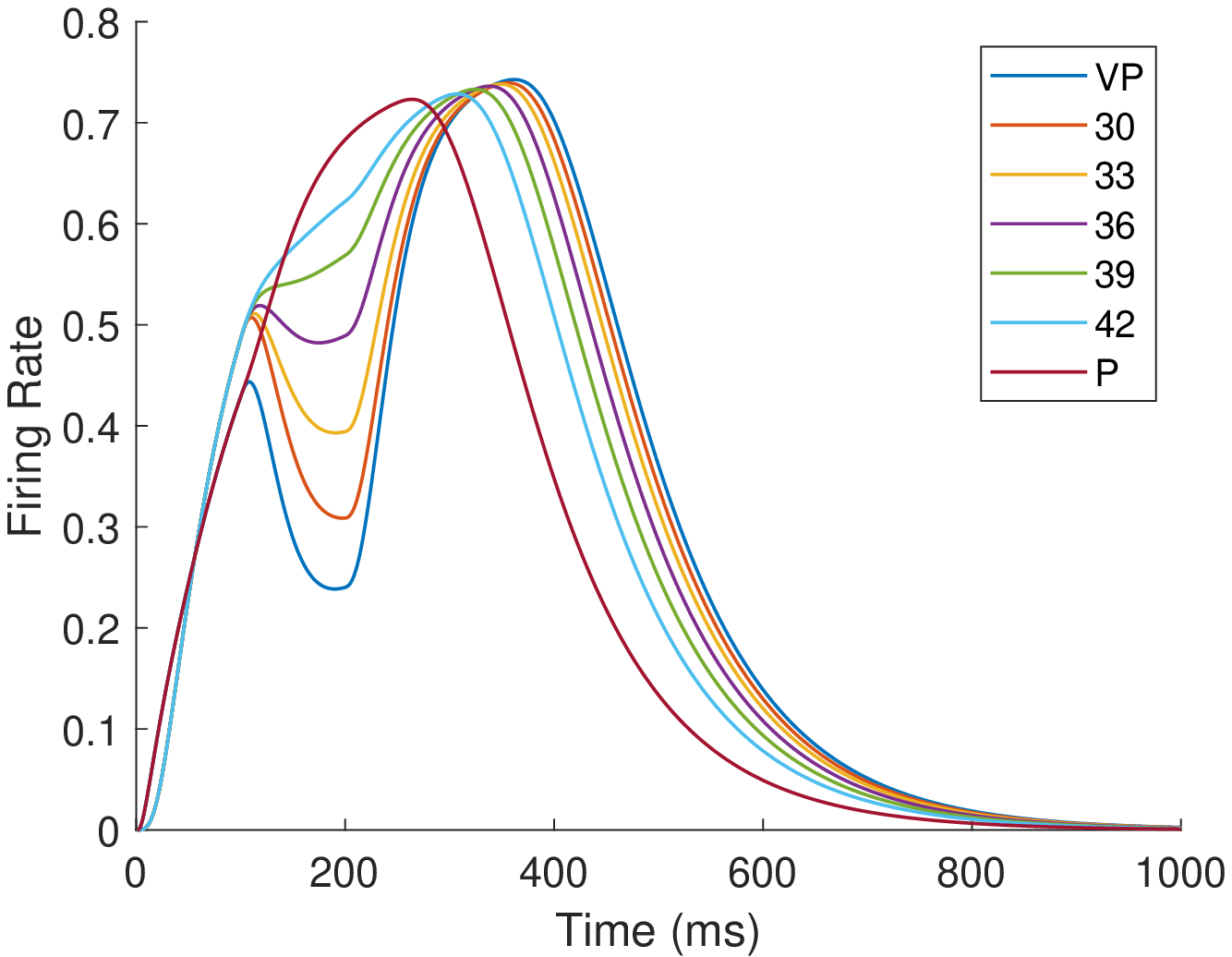}
}
\caption{The dynamic changes of neuronal firing rates in TPJ and AI areas in different tasks.}
\label{AN}
\end{figure}

In our computational model, neurons in the TPJ and AI areas exhibit similar properties. Figure \ref{AN} shows the dynamic changes of neuronal firing rates in TPJ and AI areas in different tasks. In order to make the comparison of results more intuitive and significant, the results of visual-proprioceptive congruent (VP) condition, proprioception-only (P) condition and part of visual-proprioceptive conflict (VPC) conditions are selected for display. The VPC conditions includes the results when proprioception is $0^{\circ}$ and visual deflection is $30^{\circ}, 33^{\circ}, 36^{\circ}, 39^{\circ}, 42^{\circ}$, respectively. Clearly, neurons in the TPJ area consistently exhibit integration properties. With the increase of visual disparity, the integration intensity of neurons in TPJ becomes weaker and the firing rate of neurons becomes lower. Neurons in the AI region exhibit the properties of separation and integration at different stages. In the early stage (such as about 200ms), the neurons in the AI area exhibit separation properties, and the firing rate of neurons increases with the increase of visual disparity. In the later stage (such as about 400ms), the neurons in the AI area exhibit integration properties, and the firing rate of neurons decreases with the increase of visual disparity. 

The reason for the integration effect in TPJ is that when the visual disparity is small, the receptive field overlap of proprioceptive information and visual information is huge, so the integration effect is strong and the firing rate of neurons is high; when the visual disparity is huge, the receptive field overlap of proprioceptive information and visual information is small, so the integration effect is weak and the firing rate of neurons is low.

AI is a high-level area of multisensory integration that receives excitatory stimulus (proprioc-eption-visual integration information) from TPJ, and inhibitory stimulus from S1 (proprioception information) and EBA (visual information) simultaneously. 
In the separation stage, the firing rate of neurons is highest in the proprioception-only task and the lowest in the visual-proprioceptive congruent task, and in the visual-proprioceptive conflict task, the firing rates of the neurons increased with the increase of visual disparity. The reason is that when the proprioceptive stimulation disappears, the firing rate of neurons in S1 decreases; when the visual stimulation appears, the firing rate of neurons in EBA increases, and the inhibitory effect of information from EBA on information from TPJ in AI increases. In the proprioception-only task, the firing rate of neurons in AI is highest due to the absence of inhibition from EBA. In the visual-proprioceptive congruent task, the receptive field of proprioceptive information and visual information overlap completely, the inhibitory effect of EBA on TPJ was strongest in AI, so the firing rate of neurons in AI is the lowest. In the visual-proprioceptive conflict task, the information integration in TPJ mainly occurs at the location where the proprioceptive information and visual information receptive fields overlap. With the increase of visual disparity, the inhibition of EBA on the integration information in TPJ decreases, leading to the increase of the firing rate of neurons in AI. In the integration stage, the proprioceptive and visual stimulation disappears, and their firing rates decrease. Due to the cumulative effect, the firing rate of neurons in TPJ continues to increase, and the inhibition of S1 and EBA in AI on TPJ decreases, highlighting the integration effect of TPJ.

\section{Materials and Methods}

The brain-inspired bodily self-perception model is shown in Figure \ref{model}.

\begin{figure}[htbp]
\centering
\subfigure[Firing rate of the neuron]{
\includegraphics[width=2.9in]{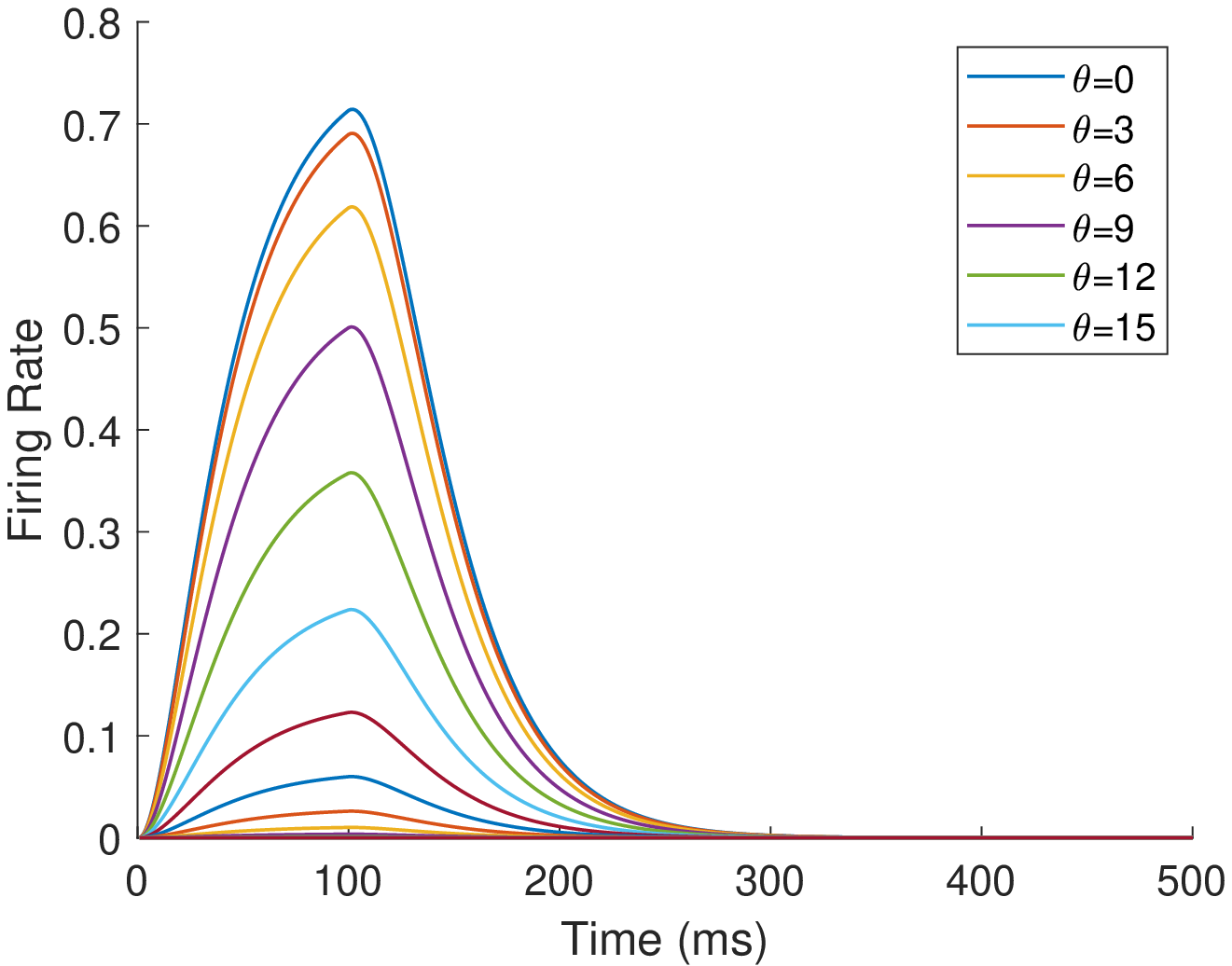}
}
\quad
\subfigure[Receptive fields of different neurons]{
\includegraphics[width=2.9in]{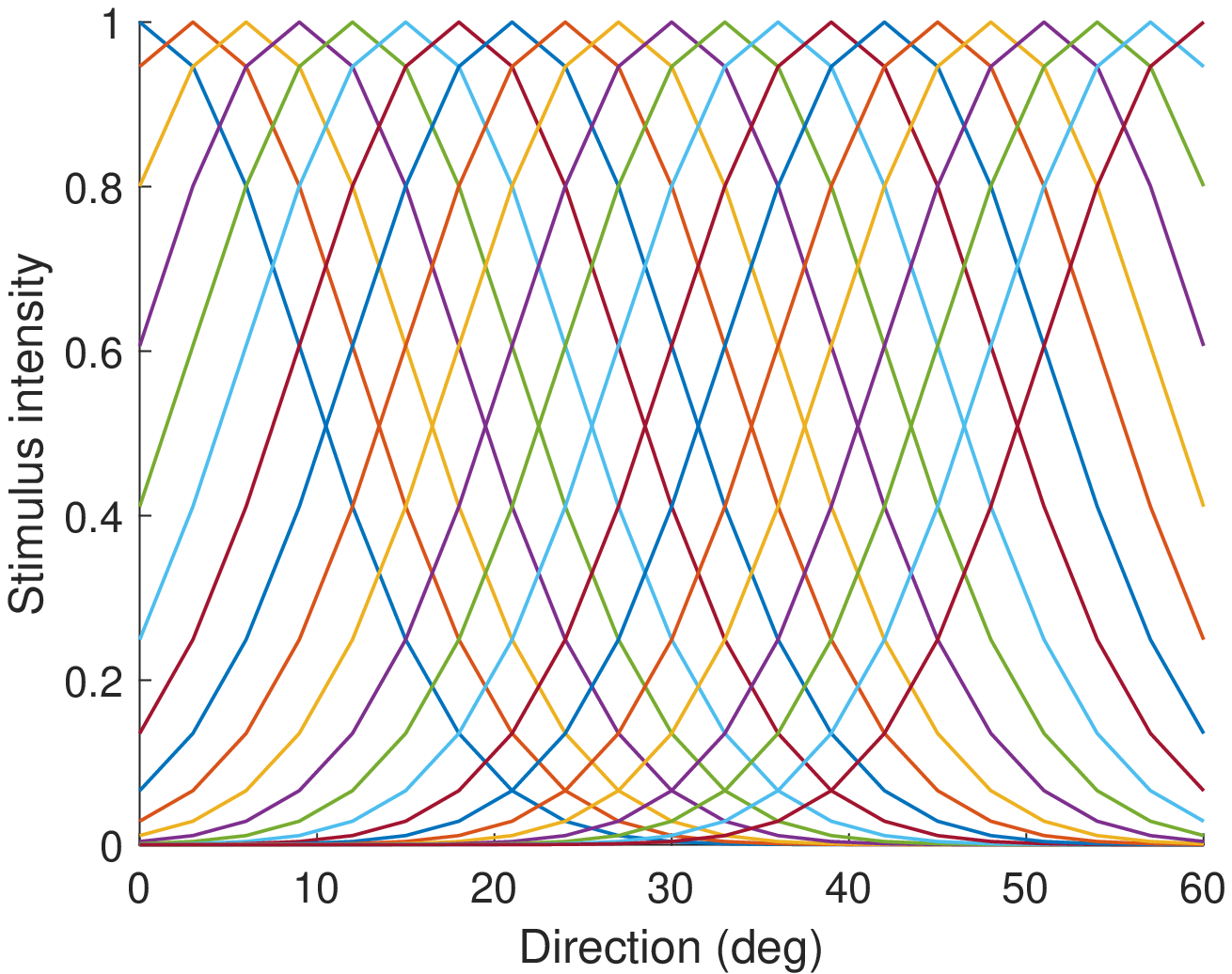}
}
\caption{The characteristic behaviors of neurons. \textbf{(a)} The firing rate of the neuron that represents $0^{\circ}$ when receiving $0^{\circ}$ stimulation, the stimulation is presented at 0 ms and ends at 100 ms.\textbf{(b)} Receptive fields of different neurons.}
\label{neuron}
\end{figure}

The neurons representing angles in M1 and V are described by Equation~\ref{eqNeuron}. $S_{j}(t)$ represents stimulus j at time $t$, and V$_{j}$(t) represents the firing rate of the neuron. The intensity of the stimulus is related to the receptive field of the neuron. Equation~\ref{eqNeuronF} describes the receptive field of the $\theta_{j}$. When the stimulation angle is $J$, the stimulation intensity of neuron representing $J$ is the largest, and the stimulation intensity of other neurons decreases successively. The firing rate of the neuron will increase when the stimulus is present and will decay when the stimulus has ended. $C$ is the parameter that controls the rate of increase and decrease. We set $C=0.04$ in M1, V, S1 and EBA areas, $C=0.01$ in TPJ area, and $C=0.15$ in AI area. The firing rate of the neuron and the receptive fields of different neurons are shown as Figure \ref{neuron}.

\begin{equation}
\label{eqNeuron}
\Delta V_{j}(t)  = -C \times (V_{j}(t)-S_{j}(t))
\end{equation}

\begin{equation}
\label{eqNeuronF}
s= e^{(\theta -\theta_{J})/2\delta^2}
\end{equation}

Neurons in other areas are described by Equation~\ref{eqNeuronP}. The $V_{i}$ is the firing rate of the postsynaptic neuron, $V_{j}$ is the firing rate of the presynaptic neuron, and the $W_{ij}$ is the synaptic weight between the presynaptic neuron $j$ and the postsynaptic neuron $i$. Specifically, the neurons in each brain area in the computational model are described as shown in Equation~\ref{eqNeuronA}.

\begin{equation}
\label{eqNeuronP}
\begin{split}
& V_{i}(t+1) = V_{i}(t) + \Delta V_{i}(t) \\
& \Delta V_{i}(t) = -C \times (V_{i}(t)-tanh(W_{ij} \times V_{j}(t))) \\
\end{split}
\end{equation}

\begin{equation}
\label{eqNeuronA}
\begin{split}
& V_{S1}(t+1) = V_{S1}(t) + \Delta V_{S1}(t) \\
& \Delta V_{S1}(t) =  -C \times (V_{S1}(t)-tanh(W_{M1-S1} \times V_{M1}(t))) \\
& V_{EBA}(t+1) = V_{EBA}(t) + \Delta V_{EBA}(t) \\
& \Delta V_{EBA}(t) = -C \times (V_{EBA}(t)-tanh(W_{V-EBA} \times V_{V}(t))) \\
& V_{TPJ}(t+1) = V_{TPJ}(t) + \Delta V_{TPJ}(t) \\
& \Delta V_{TPJ}(t) = -C \times (V_{TPJ}(t)-tanh(W_{S1-TPJ} \times V_{S1}(t) + W_{EBA-TPJ} \times V_{EBA}(t))) \\
& V_{AI}(t+1) = V_{AI}(t) + \Delta V_{AI}(t) \\
& \Delta V_{AI}(t) = -C \times (V_{AI}(t)-tanh(W_{S1-AI} \times V_{S1}(t) + W_{TPJ-AI} \times V_{TPJ}(t) + W_{EBA-AI} \times V_{EBA}(t))) \\
\end{split}
\end{equation}

The synaptic plasticity in this model is defined as shown in Equation~\ref{eqW}. $W_{ij}$ in this model represents the synaptic weight between the postsynaptic neuron $i$ and the presynaptic neuron population $j$.

\begin{equation}
\label{eqW}
W_{ij}(T+1) = W_{ij}(T) + \Delta W_{ij}(T+1) \times W_{inhibit}(T+1)
\end{equation}
where
\begin{equation*}
\label{eqWd}
\Delta W_{ij}(T+1)= \int_{T}^{T+1}\Delta w_{ij}(t)dt
\end{equation*}

The $\Delta w_{ij}(t)$ is calculated using Equation~\ref{eqWs}. 

\begin{equation}
\label{eqWs}
\Delta w_{ij}(t) = \alpha V_{i}(t)V_{j}(t)+\beta V'_{i}(t)V_{j}(t)+\gamma V_{i}(t)V'_{j}(t)
\end{equation}
where
\begin{equation*}
\alpha=\int_{-\infty }^{+\infty } f(u)du, \beta=\int_{-\infty }^{0} uf(u)du, \gamma=-\int_{0}^{+\infty } uf(u)du
\end{equation*}

$f(u)$ is the STDP function reported in \cite{RN437,RN445,RN441}. According to our previous research in \cite{RN908}, we set the parameters of $\alpha = -0.0035$, $\beta = 0.35$, $\gamma = -0.55$ in this model.

$W_{inhibit}$ is the lateral inhibitory synaptic weight used to control the location of synaptic weight update, which is described by the Equation~\ref{eqIn}. $fs$ is the firing state of the postsynaptic neuron, and $fn$ is the number of firings of the postsynaptic neuron. If the firing rate of the postsynaptic neuron is greater than the threshold, the firing state of the neuron is 1, otherwise it is 0. We set the threshold as 0.7 in this model.

\begin{equation}
\label{eqIn}
W_{inhibit}(T+1)=tanh(W_{inhibit}(T)-\frac{2 \times \arccos fs}{\pi }\times e^{fn}-1)+1
\end{equation}
where
\begin{equation*}
fs=\left\{\begin{matrix}
1 & V_{i} > threshold \\ 
0 & V_{i} \le threshold
\end{matrix}\right.
\end{equation*}

The details of visual processing such as motion perception and appearance learning are shown in our previous work \cite{RN234,RN232}.



\bibliography{scibib}

\bibliographystyle{unsrt}

\section*{Acknowledgments}

\noindent \textbf{Funding: }This study was supported by the Key Research Program of Frontier Sciences, CAS (Grant No. ZDBS-LY-JSC013), the Strategic Priority Research Program of the Chinese Academy of Sciences (Grant No. XDB32070100), the Beijing Municipal Commission of Science and Technology (Grant No. Z181100001518006).

\noindent \textbf{Author contributions: } 
Conceptualization, Y.Zh. and Y.Ze.; 
Methodology, Y.Zh.; 
Software, Y.Zh. and E.L.; 
Validation, Y.Zh. and E.L.; 
Formal Analysis, Y.Zh.; 
Investigation, Y.Zh.; 
Writing - Original Draft, Y.Zh., E.L. and Y.Ze.; 
Visualization, Y.Zh.; 
Supervision, Y.Ze.; 
Project Administration, Y.Ze.; 
Funding Acquisition, Y.Ze. 

\noindent \textbf{Competing interests: }
All authors declare that there is no conflict of interest.



\section*{Supplementary materials}

\noindent \textbf{Movie}: Rubber hand illusion experiment on the iCub robot (Multisensory integration - Visual dominance, Proprioception dominance, Proprioception based)

\noindent \textbf{Description}: 

\noindent \emph{Part I}: Experiment on the iCub robot (Multisensory integration - Visual dominance): When the hand rotation angle is small (small disparity angle), the proprioceptive drift is small. The robot mainly relies on visual information for decision-making. 

\noindent \emph{Part II}: Experiment on the iCub robot (Multisensory integration - Proprioception dominance): When the hand rotation angle is medium (medium disparity angle), the proprioceptive drift is medium. The robot mainly relies on proprioceptive information for decision-making. 

\noindent \emph{Part III}: Experiment on the iCub robot (Multisensory integration - Proprioception based): When the hand rotation angle is large (large disparity angle), the proprioceptive drift is zero. The robot completely relies on the proprioceptive information for decision-making. 

\noindent \textbf{Movie Address}: \url{https://youtu.be/G941vAfaJ-Q}


\end{document}